\documentclass{emulateapj}
\usepackage{graphicx}
\usepackage{natbib}
\usepackage{apjfonts}
\usepackage{amsmath}
\usepackage{color}
\usepackage{comment}
\usepackage{rotating}

\bibpunct{(}{)}{;}{a}{}{,}

\definecolor{joerg}{rgb}{0.7, 0.4, 0.0}

\definecolor{reply}{rgb}{0.1, 0.5, 0.0}

\newcommand{\hm}{h^{-1}}
\begin{document}

\title{Studying Inter-Cluster Galaxy Filaments Through Stacking gmBCG Galaxy Cluster Pairs}
\author{Yuanyuan Zhang\altaffilmark{1}, J\"org P. Dietrich\altaffilmark{1,2,3}, Timothy A. McKay\altaffilmark{1,2,4}, Erin S. Sheldon\altaffilmark{5}, and Alex T. Q. Nguyen\altaffilmark{1,6}}
\altaffiltext{1}{Department of Physics, University of Michigan, Ann Arbor, MI, USA}
\altaffiltext{2}{Michigan Center For Theoretical Physics,  University
  of Michigan, Ann Arbor, MI, USA}
\altaffiltext{3}{Universit\"ats-Sternwarte M\"unchen, Munich, Germany}
\altaffiltext{4}{Department of Astronomy, University of Michigan, Ann Arbor, MI, USA}
\altaffiltext{5}{Brookhaven National Laboratory, Upton, NY, USA}
\altaffiltext{6}{University of Rochester School of Medicine and Dentistry, Rochester, NY, USA}
\begin{abstract}
We present a method to study the photometric properties of galaxies in filaments by stacking the galaxy populations between pairs of galaxy clusters. Using Sloan Digital Sky Survey data, this method can detect the inter-cluster filament galaxy overdensity with a significance of $\sim 5 \sigma$ out to $z=0.40$. Using this approach, we study the $g-r$ color and luminosity distribution of filament galaxies as a function of redshift. Consistent with expectation, filament galaxies are bimodal in their color distribution and contain a larger blue galaxy population than clusters. Filament galaxies are also generally fainter than cluster galaxies. More interestingly, the observed filament population seems to show redshift evolution at $0.12<z<0.40$: the blue galaxy fraction has a trend to increase at higher redshift: a filament "Butcher Oemler Effect". We test the dependence of the observed filament density on the richness of the cluster pair: richer clusters are connected by higher density filaments. We also test the spatial dependence of filament galaxy overdensity: this quantity decreases when moving away from the inter-cluster axis between a cluster pair. This method provides an economical way to probe the photometric properties of filament galaxies and should prove useful for upcoming projects like the Dark Energy Survey.
\end{abstract}

\section{Introduction}

In the standard picture of structure formation after the Big Bang,
initial density perturbations grow via gravitational instability and
form massive structures in a bottom-up, hierarchical growth scenario
in which small, gravitationally bound structure like galaxies
conglomerate into progressively larger structures like galaxy groups
and clusters. On the largest scales, gravitational instability leads
to a collapse of matter first into sheets \citep{1970A&A.....5...84Z}
and then into a network of filaments \citep{1983MNRAS.204..891K} with
galaxy clusters at the intersection of filaments. This early result of
\citeauthor{1983MNRAS.204..891K} has been confirmed in countless
$N$-body simulations since then
\citep[e.g.,][]{1985ApJ...292..371D,1991ComPh...5..164B,2005Natur.435..629S}.

Observationally the filamentary large-scale structure (LSS) has been
traced by the galaxy distribution in redshift surveys for decades;
from first indications seen by \citet{1978MNRAS.185..357J} over the
seminal work of \citet{1989Sci...246..897G} to modern redshift surveys
like 2dF \citep{2001MNRAS.328.1039C} and SDSS
\citep{2012ApJS..203...21A}. The gaseous Warm-Hot Intergalactic Medium
residing in filaments was seen in X-ray emission
\citep{2008A&A...482L..29W,2011MNRAS.415.1961F} and absorption
 \citep{2009ApJ...695.1351B,2010ApJ...714.1715F}. \citet{2013MNRAS.tmp.1162M}
 use the lobes of giant radio galaxies in an attempt to probe the
 physical condition of the WHIM below the densities accessible by
 X-ray observations. It is, however, not clear that assumption they
 make on the state of the gas in radio lobes (equipartition and
 minimum Lorentz factor) as well as the assumption on pressure
 equilibrium of the radio lobes with the surrounding WHIM are fully
 justified. More recently the underlying Dark Matter skeleton of
large-scale structure filaments was also detected through its
gravitational lensing effect
\citep{2012Natur.487..202D,2012MNRAS.426.3369J}.

Despite these observational advances, filaments remain relatively
little studied, yet they are astrophysically interesting
places: LSS filaments contain a plurality of all matter in the
Universe \citep{2010MNRAS.408.2163A} and they harbor the ``missing
baryons'' at low redshift \citep{2001ApJ...552..473D}. It is well
known that galaxies in clusters are redder than galaxies in the
surrounding field and mostly have ceased star-formation. As galaxies
are constantly accreted into clusters along filaments, it is only
reasonable to assume that a portion of the transformation from
actively star-forming galaxies to passively evolving ones happens in
filaments. The specific role of the filamentary environment in this
process has not been studied very much. An exception is, e.g., the
work of \citet{2008MNRAS.388.1152P}, which found a spike in
star-formation activity in filament galaxies at fixed cluster centric
distance. It is of course well established that the local density of
galaxies is one of the factors deciding the efficiency with which
star-formation is quenched \citep{2010ApJ...721..193P}. The exact role
of the filament versus the galaxy-group environment, however, has not
been studied in detail.

The relative dearth of papers about filaments compared to, e.g., the
number of studies done of galaxy clusters is due to the
low-density contrast of filaments with respect to the mean density of
the Universe. While galaxy clusters by definition have a density
contrast $\delta > 200$, the typical density contrast of filaments is
$\delta < 20$ \citep{2010MNRAS.408.2163A}. This makes
filaments much more difficult to find, observe, and study, and they have generally been traceable in spectroscopic surveys only
to redshift $\sim 0.2$.

 In this paper, we develop an algorithm to study the photometric properties of inter-cluster galaxy filaments from redshift 0.1 to 0.4. Because close pairs of galaxy clusters are generally connected by filaments \citep{2004MNRAS.354L..61P,2005MNRAS.359..272C}, instead of trying to detect single filaments, our approach is to identify galaxy cluster pairs and then stack the galaxy population between them. By comparing the stack of such populations to the stack of other selected galaxy fields, we can detect an overdensity of galaxies present in the connecting filaments. We go on to analyze the color and luminosity properties of these filament galaxies and to examine these for possible redshift evolution.

 The remainder of this paper is arranged as follows. In
 Section~\ref{section:methods}, we describe our algorithm and some
 tests we have conducted on it, and represent the statistical
 significance of our filament
 overdensity. Section~\ref{section:results} contains our
 results. We compute color evolution and
 luminosity function of filament galaxies, as well as their spatial
 and richness dependence. We summarize and discuss the implications of our results in Section~\ref{section:overview}. In Appendix~\ref{section:mock}, similar
 results from N-body simulation are provided for comparison.
 Throughout this paper, we assume a flat cosmology with
 $\Omega_\mathrm{m}=0.30$ and $h=0.7$. We use $z$ to denote redshift, and $\Sigma$ to
 denote the 2-d projected galaxy number density. All the error bars shown in this paper
 represent single standard deviation errors.

\section{Methods}
\label{section:methods}

\begin{figure*}
\begin{centering}
  \includegraphics[width=0.9\textwidth]{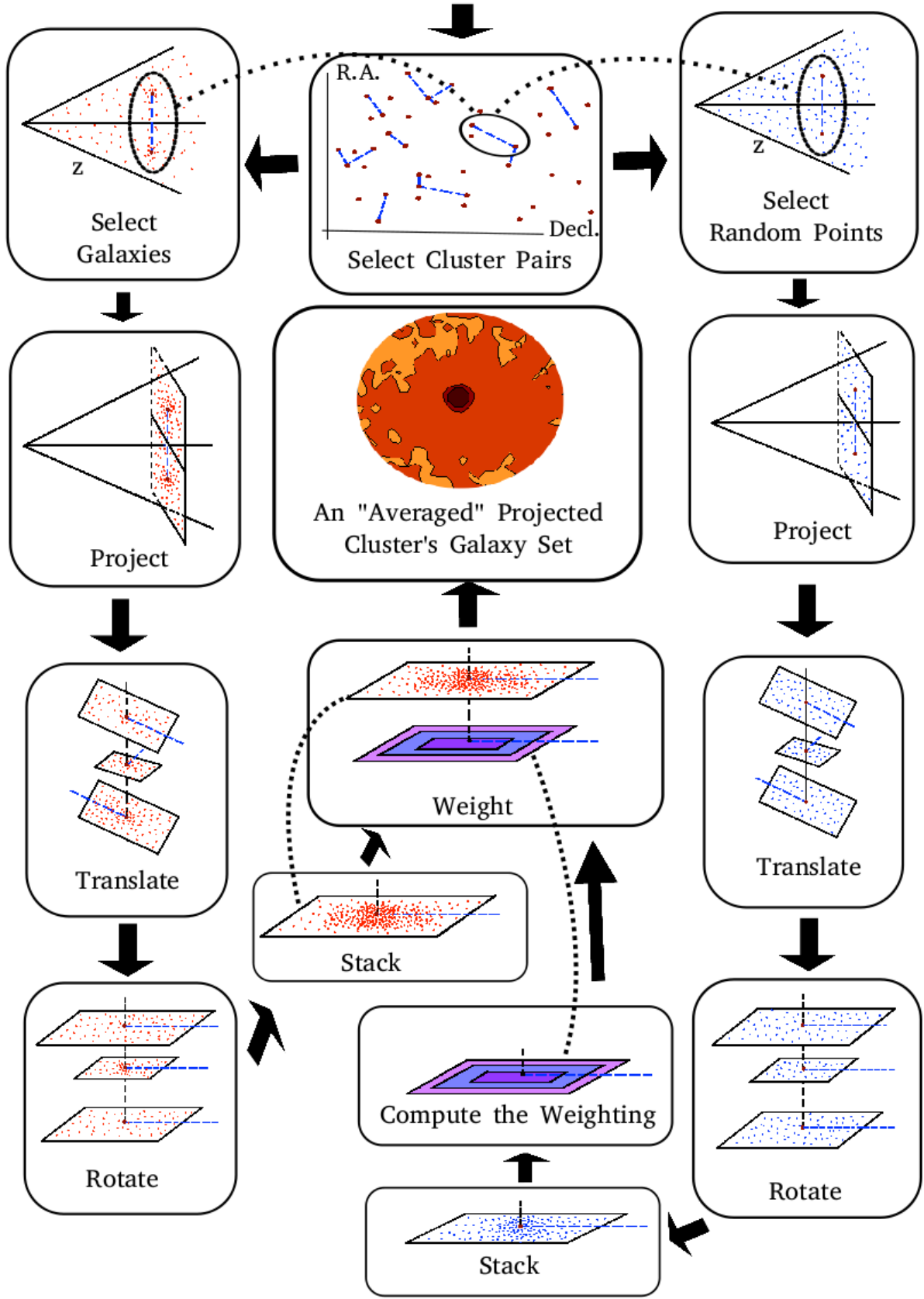}
  \caption{Flowchart illustration of the algorithm described in
    Section~\ref{section:algorithm}. The left side of the figure
    illustrates the manipulations on galaxy data in order to acquire
    stacked cluster pair galaxy fields. The right side illustrates the
    the manipulations on the random point catalog in order to acquire
    the sky coverage weighting, which tells how many times each pixel
    in the galaxy stacking is covered by real sky data. The
    manipulations on random points are identical to the manipulations
    on real galaxy data before "Compute the Weighting" step (See
    Section~\ref{section:weighting} for details). Arrows indicate the
    processing flow. We first look for cluster pairs that possibly
    have filament connections and select the galaxies that are in the
    fields of such pairs. We then re-define the coordinates of these
    galaxies with steps "Project", "Translate" and "Rotate" so that
    the coordinates reflect the galaxies' projected physical distance
    to cluster centers ( See Section~\ref{section:galaxy_processing}
    for details). We stack these galaxies and weight the stack with
    sky coverage weighting to obtain "averaged" galaxy distributions
    around one cluster center in cluster pairs. See the electronic
    edition of the journal for a color version of this figure.}
  \label{fig:algorithm}
\end{centering}
\end{figure*}

\subsection{Data}
\label{section:data}
For this study, we use the gmBCG galaxy cluster catalog \citep{2010ApJS..191..254H}. This is a large optically selected galaxy cluster catalog constructed from the SDSS Data Release 7 \citep{2009ApJS..182..543A}. It  includes 55,424 clusters in the redshift range $0.1<z<0.55$, covering the whole SDSS Data Release 7 area. When building this catalog, \citeauthor{2010ApJS..191..254H} searched for overdensities of red sequence galaxies around brightest cluster galaxy (BCG) candidates. The final catalog contains positions, estimated redshifts, and richness for each detected cluster.  Comparing to other previously well accepted cluster catalogs \citep{2005AJ....130..968M, 2007ApJ...660..239K} in the SDSS footprint, this catalog has larger sky coverage, extends to deeper redshift, and has been extensively tested by the authors of this paper.

The galaxy catalog we use for this study is constructed from SDSS DR8 BOSS imaging data \citep{2011AJ....142...72E, 2011ApJS..193...29A}. The sky coverage of this catalog is described by a catalog of random points, which samples the same angular coverage as the survey objects. Instead of using SDSS DR7, we choose to use galaxy data from the BOSS survey in a newer SDSS data release because of its improved photometry. This choice leaves the galaxy catalog having a slightly different sky coverage with the gmBCG cluster catalog. However, with the sky coverage weighting described in Section~\ref{section:weighting}, only the overlapping area of the galaxy catalog and the cluster catalog contributes to our final results.

\subsection{Algorithm}
\label{section:algorithm}

Figure~\ref{fig:algorithm} outlines our algorithm, which is described
in detail in this section. To summarize, we select pairs of galaxy
clusters and identify the galaxies
which occupy the regions around these cluster pairs. We stack these galaxies, weight for sky coverage, and compare the galaxy population between cluster pairs to the galaxy population seen in other
fields.

To test our methods and compare to our SDSS results, we also apply our algorithm to the Millennium simulation \citep{2005Natur.435..629S}. The procedures and results of this application are discussed in the appendix.

\subsubsection{Cluster Pair Selection}
\label{section:cluster_pair_selection}
\begin{figure}
\begin{centering}
  \plotone{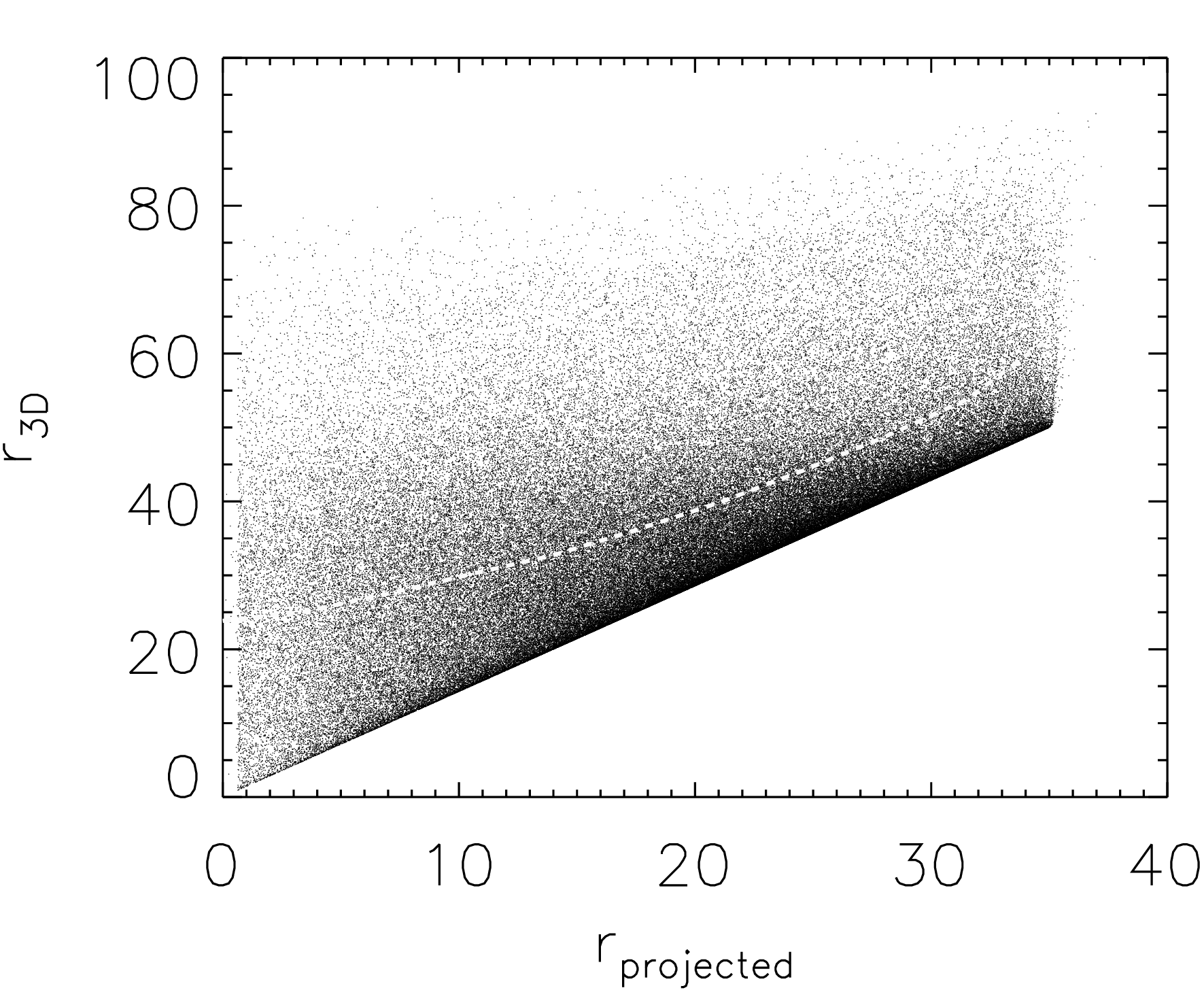}
  \caption{Cluster pairs' projected separation on the plane of the sky
    at their median redshift, $r_\mathrm{projected}$, versus their 3-d
    physical separation, $r_\mathrm{3D}$. The white dashed line shows
    the mean $r_\mathrm{3D}$ value at different
    $r_\mathrm{projected}$. We assume the clusters' photometric
    redshift are purely due to Hubble flow and use the law of cosine
    for calculating $r_\mathrm{3D}$. The cluster pairs' projected
    separation on the plane of the sky is cut off at
    $\sim35h^{-1}\mathrm{Mpc}$, depending on the exact redshift of the
    projection, but their 3-d separation can be as large as
    $90h^{-1}\mathrm{Mpc}$.}
  \label{fig:separation}
\end{centering}
\end{figure}

We begin by selecting the cluster pairs to be used in
our algorithm. According to \cite{2005MNRAS.359..272C}, the probability of finding a filament between galaxy clusters is strongly dependent on their
separation. Clusters separated by less than
$5h^{-1}\mathrm{Mpc}$ always have a connection between them, but this is primarily formed from the outer regions of the cluster populations rather than a separate filament population. For clusters that are separated between
$15h^{-1}\mathrm{Mpc}$ to $25h^{-1}\mathrm{Mpc}$, about $1/3$ of them have a filament
connection. For clusters that are separated less than
$35h^{-1}\mathrm{Mpc}$, the probability of filament existence between them is larger than $10\%$. Clusters separated more than $50h^{-1}\mathrm{Mpc}$ are unlikely to have filaments between them. Also, richer clusters tend to have more filaments connecting to them, but the number of connections rarely exceeds $5$. Informed by these expectations from simulation, we select cluster pairs separated by less than $\sim35h^{-1}\mathrm{Mpc}$ while limiting the maximum number of cluster pairs one cluster can appear in. For each cluster, the clusters paired to it satisfy the following conditions,

\begin{enumerate}
\item The difference in photometric redshift of the BCGs of the two
  clusters should differ by $<0.02$. This redshift difference roughly
  corresponds to a comoving distance of $50h^{-1}\mathrm{Mpc}$.
\item The angular separation between the two clusters is smaller than
  $35h^{-1}\mathrm{Mpc}$ on the plane of the sky at the redshift of
  the the cluster to-be-paired-to. Because we do not directly
  constrain the two clusters' 3-d physical separation, the cluster
  pairs' 3-d physical separation can be larger than
  $35h^{-1}\mathrm{Mpc}$, as shown in Figure~\ref{fig:separation}. 
\item Each cluster can appear in several cluster pairs if it has more than one neighbor meeting the above criteria. If a cluster appears in more than 5 cluster pairs, we keep its 5 closest pairs along with all pairs that include clusters with fewer than 5 neighbors.
\end{enumerate}

With this selection procedure, we identify a total of 160,954 cluster pairs. As discussed in Section~\ref{section:brightness_cut}, when stacking galaxy populations that are around these cluster pairs, we apply an absolute magnitude cut which is above the SDSS completeness limit at $z=0.38$, even with dust extinction correction and  K-correction. To avoid including galaxy populations in very dusty sky regions, we exclude cluster pairs which fall within $1^\circ \times 1^\circ$ angular boxes around galaxy objects which have dust extinction in r-band larger than 0.4. This lowers the number of cluster pairs to 158,897 in the final catalog. In this cluster pair catalog, one cluster on average appears in 5.7 cluster pairs.

\subsubsection{Selection of Galaxies}
\label{section:galaxy_set}

 Inter-cluster filaments do not always strictly lie on the inter-cluster axis. According to the studies of \citeauthor{2004MNRAS.354L..61P} and \citeauthor{2005MNRAS.359..272C}, maybe $40\%$ filaments run straightly from one cluster to the other one, and maybe a similar portion still appearing to be strand-like but curves from one cluster to the other one, and the rest being wall or sheet-like, irregular or even permeative with nondefinitive morphologies. In this paper, we do not try to distinguish filaments of different morphologies. We try to incorporate all kinds of genuine structures that cause an overdensity of matter between a cluster pair. 

For every cluster pair, we identify all galaxies in and around the two clusters. We first identify all galaxies in a rectangular box centered on the filament midpoint, with the width of the inter-cluster distance but twice as long. Galaxies in the square centered on each cluster form the "galaxy set" for that cluster. Each galaxy set includes the cluster, an inter-cluster region in which we expect to find a filament, and an opposing field region in which we do not (necessarily) expect a filament.

When selecting these neighboring galaxies, we do not make any redshift cuts. It is possible that a carefully designed photometric redshift cut could increase the significance of the results discussed in Section~\ref{section:maps}, but such a cut risks to introducing color and magnitude dependent selection effects. While photometric redshifts for bright passively evolving galaxies can be smaller than $0.05$ \citep{2009MNRAS.396.2379C}, those for fainter or star-forming galaxies can be substantially worse.

\subsubsection{Cluster Galaxy Set Processing}
\label{section:galaxy_processing}

For each cluster in one cluster pair, there is one galaxy set corresponding to it. We denote the angular coordinates of one cluster in one cluster pair to be $(\alpha_0, \delta_0)$, its comoving distance to be $d_\mathrm{C}$, the angular coordinates of the $g$th galaxy in its galaxy set to be $(\alpha_\mathrm{g}, \delta_\mathrm{g})$, and the angular coordinates of the other cluster BCG in this cluster pair to be $(\alpha_\mathrm{0p}, \delta_\mathrm{0p})$. Before stacking the galaxy sets of different clusters, we need to redefine galaxy coordinates so that they convey the same physical scale at different redshifts:
\begin{enumerate}
\item Project these galaxies onto a tangent plane at the cluster BCG,
  \begin{equation}
    \begin{split}
      \theta_\mathrm{g} & = \alpha_\mathrm{g} \cos(\delta_0),\\
      \delta_\mathrm{g} & =\delta_\mathrm{g},
      \label{eq:project}
    \end{split}
  \end{equation}
so that $1^\circ$ in the $\theta_\mathrm{g}$ and $\delta_\mathrm{g}$ dimension corresponds to the same great-circle distance.

\item Translate $(\theta_\mathrm{g}, \delta_\mathrm{g})$ as in
  \begin{equation}
    \begin{split}
      \theta_\mathrm{g}^\prime&=\theta_\mathrm{g}-(\alpha_0 \cos(\delta_0)),\\
      \delta_\mathrm{g}^\prime&=\delta_\mathrm{g}-\delta_0,
    \end{split}
  \label{eq:translate}
  \end{equation}
so that the cluster center of the galaxy set is placed at $(0, 0)$.
\item Rotate the galaxy set with a rotation matrix $M$,
  \begin{equation}
    \begin{split}
      \left[ \begin{array}{cc}
              \theta_\mathrm{g}'' \\
              \delta_\mathrm{g}''
             \end{array} \right]
&=M\times \left[ \begin{array}{cc}
              \theta_\mathrm{g}' \\
              \delta_\mathrm{g}'
             \end{array} \right]
    \end{split}
  \label{eq:rotate}
  \end{equation}
  to have the inter-cluster axis lay on $y=0$, with the direction extending to the other cluster
  aligned toward $x>0$. The rotation matrix $M$ has the form of:
  \begin{equation}
    \begin{split}
    M=\left[ \begin{array}{cc}
              \cos(\arctan \frac{\delta_\mathrm{0p}-\delta_0}{(\alpha_\mathrm{0p}-\alpha_0) \cos\delta_0}) & \sin(\arctan \frac{\delta_\mathrm{0p}-\delta_0}{(\alpha_\mathrm{0p}-\alpha_0) \cos\delta_0}) \\
             -\sin(\arctan \frac{\delta_\mathrm{0p}-\delta_0}{(\alpha_\mathrm{0p}-\alpha_0)\cos\delta_0}) & \cos(\arctan\frac{\delta_\mathrm{0p}-\delta_0}{(\alpha_\mathrm{0p}-\alpha_0)\cos\delta_0})
             \end{array} \right].
    \end{split}
  \label{eq:rotation_matrix}
  \end{equation}

\item Redefine the coordinates to be
      \begin{equation}
           \begin{split}
	    x &=\theta_\mathrm{g}''\times (d_\mathrm{C}\tan1^\circ),\\
	    y &=\delta_\mathrm{g}''\times (d_\mathrm{C}\tan 1^\circ).
	   \end{split}
       \label{eq:re-define}
      \end{equation}
  Here, $d_\mathrm{C}\tan1^\circ$ is the physical distance that corresponds to $1^\circ$ angular separation on the tangent plane at the cluster center. If all the galaxies in the galaxy set are at about the same redshift with the cluster BCG, the coordinates re-defined above would reflect the projected physical distance between these galaxies to the cluster BCG in the plane of the sky. This argument does not hold for the whole population since we do not make any redshift selection. However, this definition of coordinates should be proper for filament and cluster galaxies after foreground and background subtraction.
\end{enumerate}

\subsubsection{Stacking}
\label{section:stacking}

When stacking, we create a multidimensional galaxy stack in bins of cluster pair redshift. Each stack is four dimensional, including galaxy coordinates $(x, y)$, galaxy absolute magnitude $M_r$, and the observer frame dust extinction corrected $g-r$ color. The binning size in each dimension is chosen to be small enough not to smear out the details of $x$, $y$, $M_r$ and $g-r$ dependence, but not so small as to leave many bins unoccupied. For the $x$ and $y$ dimensions, the bin sizes are both $0.7\,\hm$\,Mpc, because of the low S/N in one spatial bin. For binning in $M_r$ and $g-r$, the bin sizes are $0.2$\,mag and $0.125$\,mag. The value of the stack at one particular bin $X_{i-1}<x\leq X_i$, $Y_{j-1}<y\leq Y_j$, $M_{r(l-1)}<M_r\leq M_{rl}$ and $(g-r)_{k-1}<(g-r)\leq (g-r)_k$ is denoted as $G_z(x_i, y_j, M_{rl}, (g-r)_k)$. To include a comparison of cluster galaxies to filament galaxies, we also build a two dimensional stack in $M_r$ and $g-r$ with galaxies around cluster centers which satisfy $\sqrt{x^2+y^2}<0.7\,\hm$\,Mpc. This stack of cluster galaxies is denoted as $CL_z(M_r, g-r)$.

\subsubsection{Stack Weighting}
\label{section:weighting}

Because the galaxy cluster pairs have different separations, and the galaxy sets of different clusters are selected based on their
cluster pair separations, the physical sizes of our cluster galaxy sets vary. Also, the cluster pairs may fall at the edge of the survey, and the galaxy sets will have irregular sky coverage. Without accounting for this, we would observe artificial galaxy density gradients in the $x$ and $y$ dimension of the stacking.  To remove such an artificial gradient, we weight $G_z(x, y, M_r, g-r)$ with a coverage weighting function $W_z(x, y)$. We first explain the physical interpretation of $W_z(x, y)$ and the weighting procedure in this and the next paragraph, and describe details on computing $W_z(x, y)$ in the rest of this section. The value of $W_z(x, y)$ at $X_{i-1}<x\leq X_i$ and $Y_{j-1}<y\leq Y_j$, denoted as $W_z(X_i, Y_j)$, tells how many cluster galaxy sets include valid data for the pixel at $X_{i-1}<x\leq X_i$ and $Y_{j-1}<y\leq Y_j$. For example, $W_z(X_i, Y_j)$=3 when there are 3 clusters' galaxy sets fully covering this pixel, or $W_z(X_i, Y_j)=\frac{10}{3}$ when there are 3 clusters' galaxy sets fully covering this pixel and another cluster's galaxy set covering 1/3 of this pixel.

We weight $G_z(x, y, M_r, g-r)$
with Equation~\ref{eq:gxyv}, and acquire a weighted  distribution, $g_z(x, y, M_r, g-r)$, the value of which at one bin, $X_{i-1}<x\leq X_i$, $Y_{j-1}<y\leq Y_j$, $M_{r(l-1)}<M_r\leq M_{rl}$ and $(g-r)_{k-1}<(g-r)\leq (g-r)_k$, is denoted as $g_z(x_i, y_j, M_{rl}, (g-r)_k)$:
\begin{equation}
\label{eq:gxyv}
g_z(x_i, y_j, M_{rl}, (g-r)_k)=\frac{G_z(x_i,y_j, M_{rl}, (g-r)_k)}{W_z(x_i,y_j)}.
\end{equation}
This weighted distribution, $g_z(x, y, M_r, g-r)$, can be interpreted as representing properties of one "averaged" cluster's galaxy set with filament connection on the $x>0$ side.

The $W_z(x, y)$ we use is built from the random point catalog
described in Section~\ref{section:data}. This random point catalog
first goes through the same algorithm with the galaxy catalog as
described in Section~\ref{section:galaxy_set} and
Section~\ref{section:galaxy_processing}, and then is stacked
together. Unlike Section~\ref{section:stacking}, in which we build a
four dimensional distribution from stacking galaxies, we can only
build a two dimensional distribution in $x$ and $y$ from stacking
random points, denoted as $R_z(x,y)$. At $X_{i-1}<x\leq X_i$ and
$Y_{j-1}<y\leq Y_j$, the value of $R_z(x, y)$, denoted as $R_z(X_i,
Y_j)$, tells how many random points there are in this bin. Because the
random point catalog covers the survey area evenly, $R_z(x, y)$ can be
used as a proxy of $W_z(x, y)$. The relation between $R_z(x, y)$ and
$W_z(x, y)$ at one particular bin is
\begin{equation}
\label{eq:rxy}
R_z(x_i, y_j)=W_z(x_i, y_j) \times C(z).
\end{equation}
Here $C(z)$ is the average number of random points per
[$h^{-1}$Mpc]$^2$ at redshift $z$, and $z$ is the median redshift
value of the clusters which have their galaxy sets or random point
sets stacked to build $G_z(x, y, M-r)$ or $R_z(x, y)$. $C(z)$ can be
computed from the comoving distance, denoted by $d_\mathrm{C}(z)$, at
redshift $z$, and the average number of random points per
$\mathrm{deg}^2$, denoted by $C$, through
\begin{equation}
\label{eq:cz}
C(z)=C\times \left[\arctan \left(\frac{0.7 h^{-1}\mathrm{Mpc}}{d_\mathrm{C}(z)}\right)\right]^2.
\end{equation}
With known values of $R_z(x, y)$, and $C$, We compute $W_z(x, y)$ from
Equation~\ref{eq:rxy} and Equation~\ref{eq:cz}, and then $g_z(x, y, M_r,
g-r)$ from Equation~\ref{eq:gxyv}, which is used for deriving the
scientific results in this paper.

To weight $CL_z(M_r, g-r)$, which is used for comparison with $g_z(x, y,M_r, g-r)$, we record the total number of random points that fall
around cluster centers with $\sqrt{x^2+y^2}<0.7h^{-1}$Mpc, denoted by $R_z$, and
compute the weighted function of $CL_z(M_r,g-r)$, which is denoted
by $cl_z(M_r,g-r)$, with following equations:

\begin{equation}
  \label{eq:clweight}
  \begin{split}
    R_z&=W_z\times C(z),\\
    cl_z(M_{rl}, (g-r)_k)&=\frac{CL_z(M_{rl},(g-r)_k)}{W_z}.
  \end{split}
\end{equation}

In this weighting procedure, one might worry that the poisson noise of
the random point sampling would lower the S/N of $g(x, y, M_r, g-r)$
or $cl_z(M_r, g-r)$. Here, we show that with proper random point
density, this influence is insignificant. The poisson noise in $R_z(x,
y )$ and $R_z$ increases with redshifts because $C(z)$ decreases with
redshifts. Our random point catalog samples the survey coverage with
$C=9,275$ and at $z=0.40$, $C(z)=5.98$. Since we always stack galaxy
sets of more than 5,000 clusters in each redshift bin, the S/N of
$R_z(x, y)$ at most spatial bins, and of $R_z$, is larger than
$100$. At this significance level, the weighting constructed from
$R_z(x, y)$ and $R_z$ brings in negligible noise to $g(x, y, M_r,
g-r)$ or $cl_z(M_r, g-r)$.

\subsubsection{Galaxy Absolute Magnitude and Luminosity Cut}
\label{section:brightness_cut}

For every galaxy selected in Section~\ref{section:galaxy_set}, we compute an absolute magnitude from its apparent magnitude, $m_r$, SDSS dust extinction correction, $e$,  K-correction, $K$, and its cluster BCG's luminosity distance $d_\mathrm{L}$, with equation
\begin{equation}
\label{eq:mr}
M_r = m_r - 5 \log\left(\frac{d_\mathrm{L}}{10\,\mathrm{pc}}\right) - e - K \;.
\end{equation}
We compute the K-correction for each galaxy with analytical approximations provided by \cite{2010MNRAS.405.1409C}, since the more popular SED template fitting method is inefficient to implement with our large volume of data. This \citeauthor{2010MNRAS.405.1409C} approach approximates K-corrections computed by \cite{1997A&A...326..950F} and \cite{2007AJ....133..734B} with analytical
polynomials of two parameters: redshift and one observed color, and
the residual between values measured by this approach and others \citep{1997A&A...326..950F, 2007AJ....133..734B, 2009MNRAS.398.1549R} is close to zero for r-band in the
redshift range $[0.1, 0.45]$ \citep{2010MNRAS.405.1409C}. In our application of this method, we use
each galaxies own $g-r$ color and the cluster BCG's photometric
redshift. 

Because SDSS is an apparent magnitude limited survey \citep{2002AJ....123..485S}, its imaging data contains different galaxy populations at different redshifts: at lower redshift, the galaxy catalog is more complete, including more intrinsically faint galaxies than at higher redshift. To ensure we are selecting the same galaxy population for comparison in different redshift bins, we apply a $-24.2<M_r<-20.4$ cut in Section~\ref{section:maps}, Section~\ref{section:fake}, Section~\ref{section:color}, Section~\ref{section:richness} and Section~\ref{section:spatial}. Under this magnitude cut, even with a maximum dust extinction correction, $e_\mathrm{max}=0.4$, and a maximum K-correction $K_\mathrm{max}=0.6$ in the galaxy catalog, a galaxy's absolute magnitude is still above the $95\%$ completeness limit of r-band at redshift $z=0.38$. In Section~\ref{section:luminosity}, the $M_r$ cut is different for different redshift bins, which is listed in Table~\ref{tbl:lf}. This cut ensures that the galaxy population is more than $95\%$ complete in each redshift bin.

\begin{figure*}

  \includegraphics[width=1.0\textwidth]{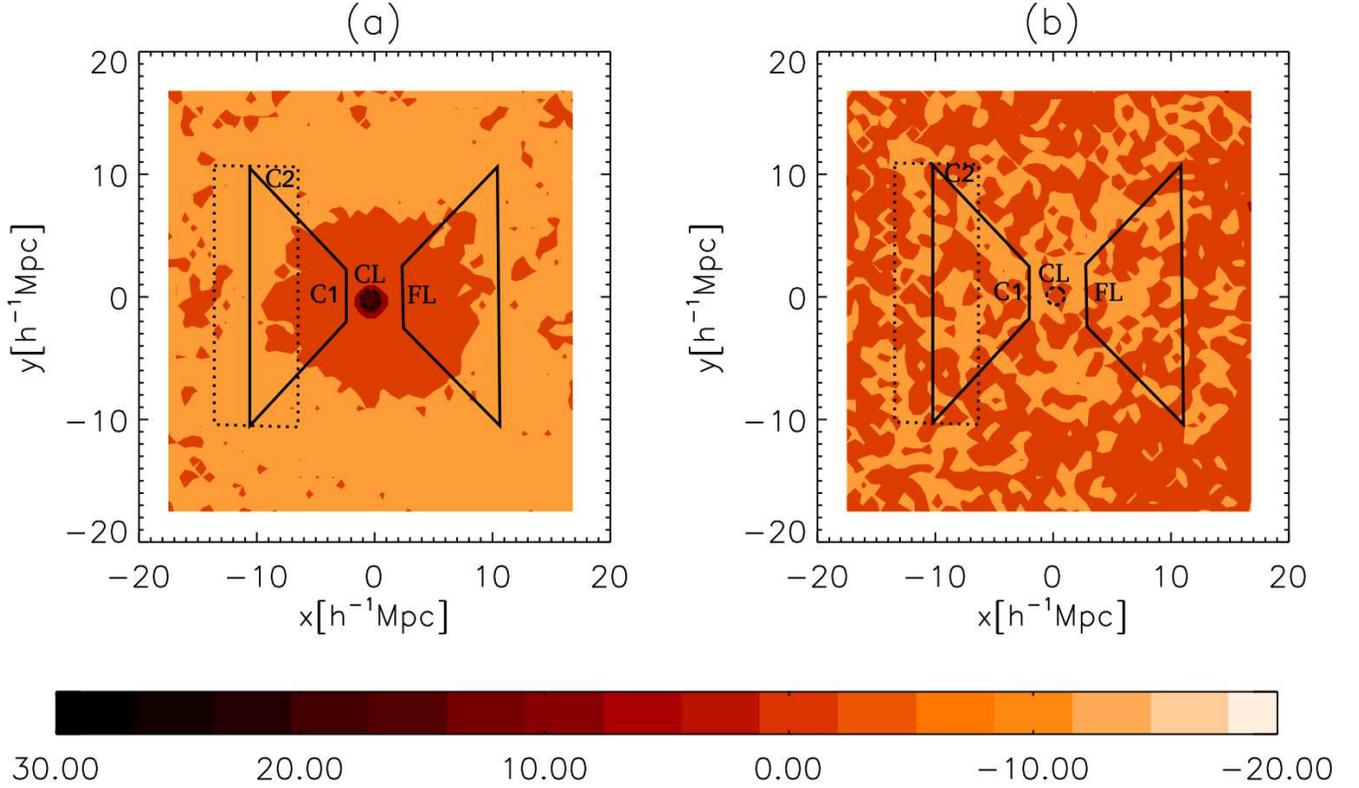}

  \caption{(a) Shows a galaxy overdensity significance contour map from stacking galaxy sets of clusters at $0.14<z<0.18$. There is a significant overdensity in the middle of the map, caused by the presence of many cluster galaxies. (b) Shows null test results from stacking randomly re-positioned pairs, where the whole field is noisily flat. In (a) and (b), the solid line and dashed line boxes mark out the four regions defined in Section~\ref{section:maps}. See the electronic edition of the journal for a color version of this figure.}
  \label{fig:contour}
\end{figure*}

\begin{figure*}

  \plotone{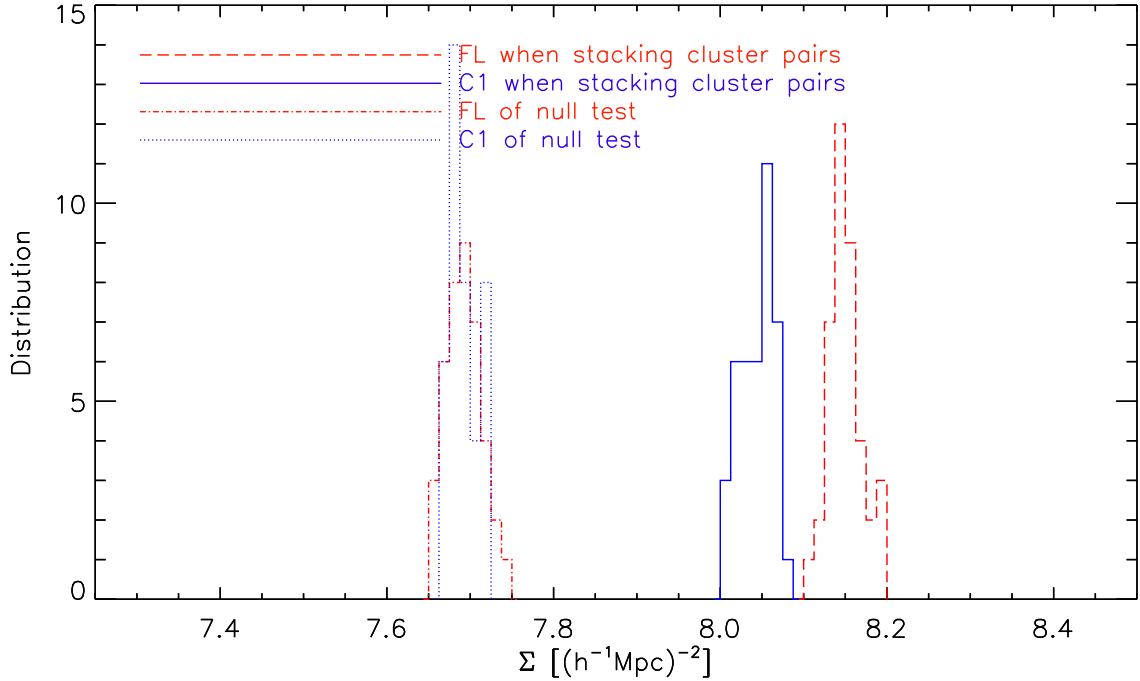}

  \caption{Projected galaxy number density in the filament region (FL) and the comparison 1 region (C1) at $0.14<z<0.18$ when stacking real cluster pairs or when performing null test with randomly re-positioned pairs (see Section \ref{section:fake} for details). The histograms shows distributions of 40 bootstrapped resamplings of the original stacking. When stacking galaxy sets of real cluster pairs, the FL region (red dashed line) displays a galaxy overdensity above the C1 region (blue solid line) with a significance of $\sim5\sigma$. Galaxy number counts of FL and C1 at other redshift slices also show similar high significance detection. In null test of stacking randomly re-positioned pairs, we don't observe any overdensity in F1 (blue dotted line) over C1 (red dash dot line). Because the galaxy overdensity in cluster fields, the galaxy count in F1 and C1 from stacking real cluster pairs are higher than stacking randomly re-positioned pairs. See the electronic edition of the journal for a color version of this figure.}
  \label{fig:bsstat}
\end{figure*}

\begin{figure*}
\begin{center}
  \plotone{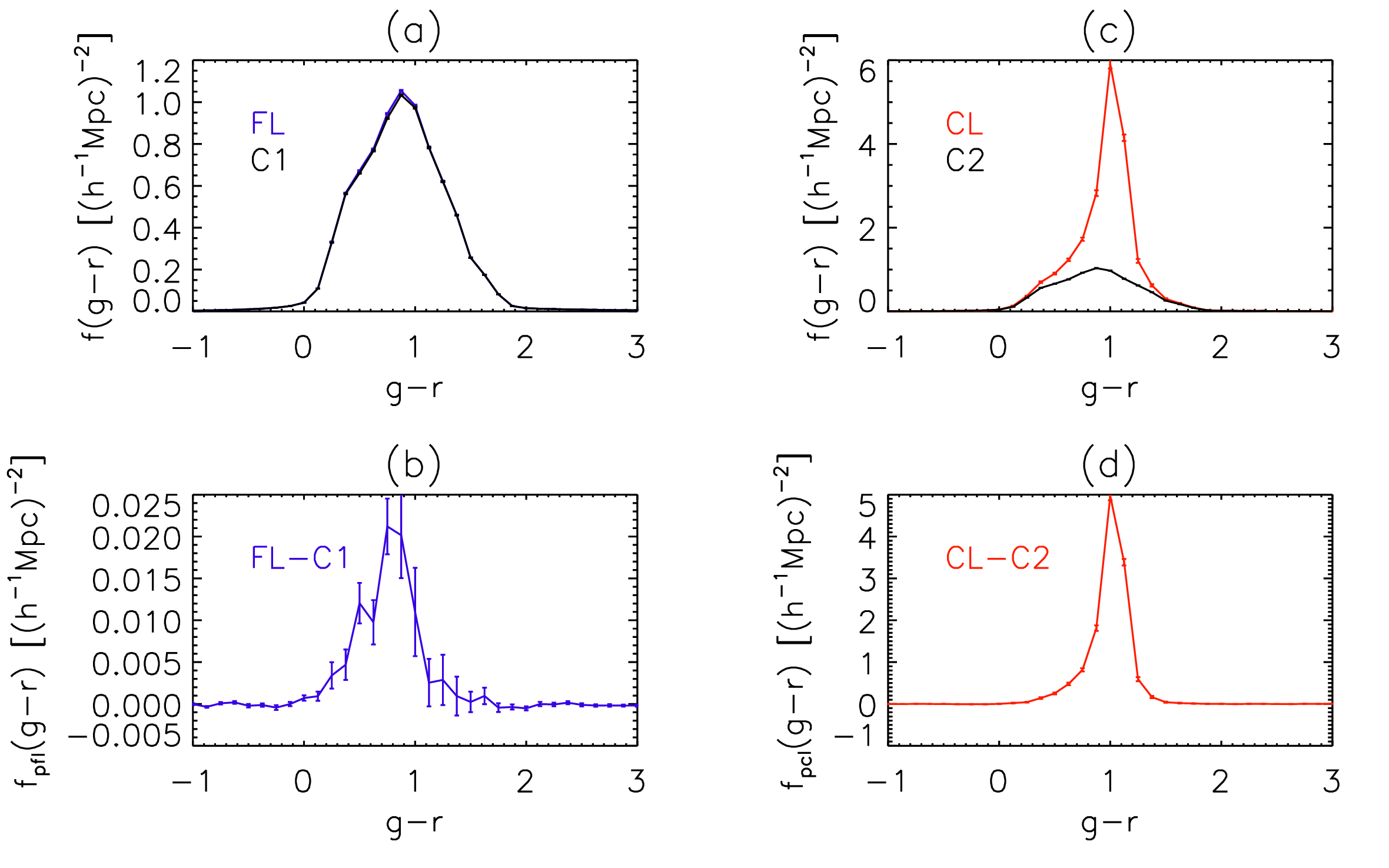}
  \caption{How we get the "pure" color distributions of filament galaxies and cluster galaxies. (a) shows the color distributions of the FL region and the C1 region, which are almost indistinguishable. After subtracting the C1 region distribution from the FL region distribution, we obtain the color distribution of pure filament galaxies, which is shown in (b). Subtraction between the CL region and the C2 region is also applied to get the pure color distribution of cluster galaxies. (c) shows the color distributions of the CL region galaxies and the C2 region, and (d) shows the pure color distribution of cluster galaxies after subtraction. These plots are based on stacking galaxy sets of clusters in the redshift bin $[0.14, 0.18]$. Unless otherwise noted, the error bars in this figure and the rest of the paper are single standard deviation errors estimated from bootstrapping the galaxy sets being stacked. Because one cluster on average appears 5.7 times in the stack (158,897 cluster pairs versus 55,424 clusters), and the error for clusters tend to be underestimated, we enlarge the error estimation of clusters by a factor of $\sqrt{5.7}$ through out the paper. See the electronic edition of the journal for a color version of this figure.}
  \label{fig:gr_ill}
\end{center}
\end{figure*}
\subsection{Foreground/Background Subtraction and Filament Detection Significance}
\label{section:maps}
In this section, we describe our foreground and background galaxy subtraction and quantify the detection significance of our filament signal. Within the weighted average cluster galaxy set $g_z(x, y, M_r,g-r)$ we define a "filament" region (denoted FL), a "cluster" region (denoted CL), and two comparison regions C1 and C2. All of these regions are displayed in Figure~\ref{fig:contour}. We search for filament galaxies by subtracting the galaxy population observed in C1 or C2 from that seen in FL.

Since the filament fields are placed on the right side of the galaxy sets, the "filament" region is defined with the following constraints:
\begin{equation*}
\begin{split}
  2.1\,\hm\,\mathrm{Mpc}<x<&10.5\,\hm\,\mathrm{Mpc} \\
  |\theta=\arctan(y/x)|&\le 45^{\circ}.
\end{split}
\end{equation*}
We define this region to be $2.1\,\hm \, \mathrm{Mpc}$ away from the cluster center
to avoid including contamination by cluster galaxies. Although
cluster galaxy contents can be removed with the foreground/background subtraction procedure as described later in this section,
contamination from this population would lower the S/N of filament
signals. We also eliminate galaxies that are too far away from cluster
centers (with $x>10.5h^{-1}\mathrm{Mpc}$), to avoid lowering the $S/N$ of filament detections with too noisy spatial bins. Since few cluster pairs extend to $10.5h^{-1}\mathrm{Mpc}$, at the spatial bins beyond this distance, not only the galaxy count noise becomes significant, the random point count noise also starts to influence, resulting in exceptionally low $S/N$ of these bins. The filament region angular extent $|\theta|=45^{\circ}$
is chosen based on discussion in Section~\ref{section:spatial}. Defining the
cluster region is more straightforward. Since cluster BCGs are placed
at $(x, y)=(0, 0)$, we define the region enclosed within
$\sqrt{x^2+y^2}<0.7h^{-1}\mathrm{Mpc}$ to be the cluster region, which is also
the coverage of functions $CL_z(M_r, g-r)$ and $cl_z(M_r, g-r)$.

To evaluate signals from the foreground/background galaxy population, we define two comparison regions. The first, C1, satisfies these constraints:
\begin{equation*}
\begin{split}
-10.5h^{-1}\mathrm{Mpc}<x&<-2.1h^{-1}\mathrm{Mpc}\\
|\theta=\arctan(y/x)|&\le 45^{\circ},
\end{split}
\end{equation*}
The second comparison region, C2, satisfies,
\begin{equation*}
\begin{split}
-13.3h^{-1}\mathrm{Mpc}<x&<-6.3h^{-1}\mathrm{Mpc}  \\
-10.5h^{-1}\mathrm{Mpc}<y&<10.5h^{-1}\mathrm{Mpc}.
\end{split}
\end{equation*}
The C1 region is symmetrical to the FL region, and is used for filament foreground/background subtraction. Both FL and C1 contain galaxies present in the outskirts of the clusters, so using C1 for foreground/background subtraction should, to first order, eliminate these. The C2 region is used for foreground/background signal evaluation in estimating the cluster population. It is picked to be a region away from the cluster center.

 In Figure~\ref{fig:bsstat}, we resample with replacement, i.e. bootstrap, cluster pairs which have their galaxy sets stacked, and show the projected galaxy number density in the FL region and the C1 region. The averaged galaxy number density in the filament region is $\sim5 \sigma$ higher than that of the C1 region. Since cluster outskirt galaxies have already been taken into consideration when making comparison, this galaxy overdensity is unlikely to be caused by the existence of one galaxy cluster alone, but related to cluster pair structures exclusively. We associate this galaxy overdensity with the filament population between cluster pairs. Such a galaxy overdensity is very small comparing to galaxy overdensity caused by cluster fields. In Figure~\ref{fig:contour}(a), we show projected galaxy overdensity significance contour map of stacked galaxy sets at $0.14<z<0.18$. At $(X_i, Y_j)$, the projected galaxy overdensity significance, denoted by $S(X_i, Y_j)$, is defined as,
\begin{equation}
\label{eq:sig}
\begin{split}
S_z(\;X_i, \;Y_j) &=\frac{\sum\limits_{k, l} g_z(\;X_i, \;Y_j, \;(M_r)_k, \;(g-r)_l)-\mu_z}{\sigma_z}.
\end{split}
\end{equation}
Here, $\mu_z$ and $\sigma_z$ are the mean and scatter of projected galaxy count per spatial pixel in the C2 region, computed with,
\begin{equation}
\begin{split}
\mu_z&=\frac{\sum\limits_{(i,j)\; \mathrm{in}\; \mathrm{C2} \;}\sum\limits_{k, l} g_z(\;X_i, \;Y_j, \;(M_r)_k, \;(g-r)_l)}{\sum\limits_{(i,j)\; \mathrm{in} \;\mathrm{ C2}}}, \\
\sigma_z &=\sqrt{\frac{\sum\limits_{(i,j)\; \mathrm{in} \; \mathrm{C2} \;}[\sum\limits_{k, l} g_z(\;X_i,\; Y_j,\; (M_r)_k, \;(g-r)_l)-\mu_z]^2}{\sum\limits_{(i,j)\; \mathrm{in} \; \mathrm{C2}}}}.
\end{split}
\label{eq:2}
\end{equation}
In Figure~\ref{fig:contour}(a), We observe a significant overdensity in the center because of cluster galaxy population. The filament overdensity is not directly observable in this figure, due to the very small density contrast of the filaments.

Because more than $90\%$ of the galaxies observed in the FL region come from foreground and background populations, the properties of filament galaxies would not be distinguishable without foreground/background subtraction. To get the pure properties of galaxy filaments, like color distribution and luminosity distribution, we first get such distributions of the total galaxy population in the FL region and the C1 region, and then subtract the C1 region count from the FL region count. For example, to get the color distribution of pure filament galaxies at one redshift bin, we first get color distribution of the FL region with
\begin{equation}
  \label{eq:fil}
f_\mathrm{FL}(g-r)=\frac{\sum\limits_{(i,j)\; \mathrm{in} \; \mathrm{FL} \;}\sum\limits_{k} g_z(X_i,Y_j,(M_r)_k, (g-r))}{\sum\limits_{(i,j)\; \mathrm{in} \; \mathrm{FL}}0.49(h^{-1}\mathrm{Mpc})^2},
\end{equation}
and color distribution of the C1 region with
\begin{equation}
  \label{eq:cmp}
f_\mathrm{C1}(g-r)=\frac{\sum\limits_{(i,j)\; \mathrm{in} \; \mathrm{C1} \;}\sum\limits_{k} g_z(X_i, Y_j, (M_r)_k, (g-r))}{\sum\limits_{(i,j)\; \mathrm{in} \; \mathrm{C1}}0.49(h^{-1}\mathrm{Mpc})^2},
\end{equation}
and then subtract $f_\mathrm{C1}(g-r)$ from $f_\mathrm{FL}(g-r)$ as in,
\begin{equation}
\label{eq:pure}
f_\mathrm{pfl}(g-r)=f_\mathrm{FL}(g-r)-f_\mathrm{C1}(g-r).
\end{equation}
$f_\mathrm{pfl}(g-r)$ is the color distribution of pure filament galaxies after removing foreground/background galaxy contents. In Equation~\ref{eq:fil} and Equation~\ref{eq:cmp}, the $0.49(h^{-1}\mathrm{Mpc})^2$ factor is in the denominators because one pixel in the $x$ and $y$ dimension of $g(x, y, M_r, g-r)$ is $0.7h^{-1}\mathrm{Mpc}\times 0.7h^{-1}\mathrm{Mpc}$. This subtraction is also illustrated in Figure~\ref{fig:gr_ill}. To acquire properties of cluster galaxies for comparison with filament galaxies, similar subtraction between the CL region and the C2 region is applied:

\begin{equation}
\label{eq:cl}
\begin{split}
f_\mathrm{CL}(g-r)&=\frac{\sum\limits_{k} cl_z( (M_r)_k, (g-r))}{0.49(h^{-1}\mathrm{Mpc})^2},\\
f_\mathrm{C2}(g-r)&=\frac{\sum\limits_{(i,j)\; \mathrm{in} \; \mathrm{C2} \;}\sum\limits_{k} g_z(X_i,Y_j,(M_r)_k,(g-r))}{\sum\limits_{(i,j)\; \mathrm{in} \; \mathrm{C2}}0.49(h^{-1}\mathrm{Mpc})^2}, \\
f_\mathrm{pcl}(g-r)&=f_\mathrm{CL}(g-r)-f_\mathrm{C2}(g-r).
\end{split}
\end{equation}
Here, $f_\mathrm{pcl}(g-r)$ is the color distribution of pure cluster galaxies.

\begin{figure*}
\begin{center}
  \plotone{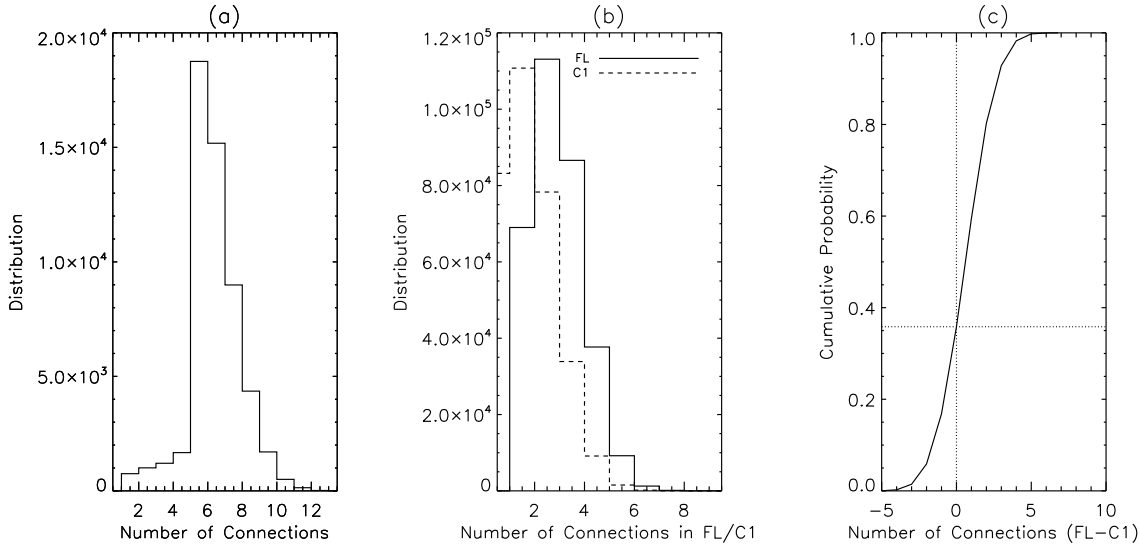}
  \caption{(a) Distribution of the number of pair connections each cluster have in the cluster pair catalog. Most clusters are connected to more than 5 other clusters. (b) Distribution of the number of cluster pair connections in the C1 region (dashed line) and the FL region (solid line). There is no cluster pair connection in C1 for $30\%$ of the time, and the average number of cluster pair connections in C1 is lower than that in FL. Meanwhile, the FL region always contain at least one cluster pair connection since one cluster pair is always aligned along $y=0$ at the FL side. (c) Cumulative probability distribution of the difference between number of cluster pair connections in FL and C1. C1 is unlikely to have equal or more cluster pairs than FL ($\sim35\%$). The horizontal and vertical dotted lines indicate where number of cluster pair connections in FL and C1 are equal.} 
  \label{fig:c1c2}
\end{center}
\end{figure*}

\subsection{Justification on Using C1/C2 for Foreground/Background Subtraction}
\label{section:justification}

When using C1/C2 for foreground/background subtraction, a concern arises that C1/C2 not only contain foreground/background galaxies, but also filaments which are not in the FL region. In this section, we show that such scenarios do exist but the detection of filament galaxies is still effective.

Indeed, more often than not, galaxy filaments also appear in regions other than the FL fields because each cluster is typically connected to more than one neighbor. In Figure~\ref{fig:c1c2}(a), we show the distribution of pair connections for each cluster in our cluster pair catalog. Most clusters have more than 5 pair connections, and some of them even have more than 10 pair connections. However, the probability of one cluster having a filament connection in C1 region is relatively low. In Figure~\ref{fig:c1c2}(b), we show the distribution of cluster pair connections appeared in C1 and FL while placing one cluster pair on the $y=0$ axis in the FL region. About $30\%$ of the clusters do not have cluster pair connection appearing in C1, and the number of cluster pair connections in C1 tend to be smaller than this number in FL. In Figure~\ref{fig:c1c2}(c), we show the cumulative probability distribution of the difference between number of cluster pair connections in FL and C1. The probability of C1 having more or or equal number of cluster pair connections than the FL region is less than $40\%$. On average, the FL region contains $1.091 \pm 0.007$ more cluster pair connections than the C1 region. Therefore, the FL region always tend to contain more filament galaxies than the C1 region, and using C1 for foreground/background subtraction would still leave proper galaxy counts similar to stacking pure inter-cluster galaxies.

Similarly to C1, the C2 region would also contain cluster-pair field galaxies. Using C2 for cluster  foreground/background subtraction will then remove cluster-pair field galaxy count in addition to foreground and background galaxy count. However, because of the very high galaxy overdensity in clusters, the absence of such a population have negligible influence ($\sim 1\%$) and won't change the main conclusions of this paper.

 Another concern about our method of searching for filaments between cluster pairs is that since clusters tend to cluster, the filament signal we are seeing might come from clusters which cluster between cluster pairs. We compare the number of clusters two cluster finders, gmBCG \citep{2010ApJS..191..254H} and redMaPPer \citep{2013arXiv1303.3562R},  find around cluster pairs and between cluster pairs. We notice that although clusters might have slightly higher chance to appear between cluster pairs than appearing randomly around them, this clustering of clusters effect contributes to $<5\%$ of the filament signal we observe.

\subsection{Null Test with Random Cluster Pair Re-Positioning}
\label{section:fake}

To test the robustness of our algorithm, we randomly translate the angular coordinates of the cluster pairs and rerun our algorithm. We expect to get a stack of galaxies which are completely flat with no overdensity either at the "cluster center" or in the "filament region".

\subsubsection{Random Cluster Pair Re-Positioning}

For one cluster pair in the 160,954 pre-masked cluster pair catalog, we first generate a random point inside the angular region $110^{\circ}<\alpha<260^{\circ}$ and $0^{\circ}<\delta<60^{\circ}$. We then translate the angular coordinates of the two clusters in this cluster pair so that the left cluster is laid on the generated random point. After randomly re-position every cluster pair in the 160,954 pre-masked cluster pair catalog, we mask out these pairs which fall into regions with dust extinction in r-band larger than 0.4, and make a new pair catalog which will be referred as the randomly re-positioned pair catalog.This new pair catalog and the original cluster pair catalog overlaps significantly in sky coverage, and have similar physical separation distributions, as well as similar cluster pair count in each redshift bin.

\subsubsection{Galaxy Set Profile of Randomly Re-positioned Pairs}

We run the algorithm described in Section~\ref{section:algorithm} and Section~\ref{section:maps} with the randomly re-positioned pair catalog, and get the $S(x, y)$ contour map as shown in Figure~\ref{fig:contour}(b), and the distributions of galaxy number count per $[h^{-1}\mathrm{Mpc}]^2$ in the FL region and the C1 region as shown in  Figure~\ref{fig:bsstat}. We do not observe an overdensity either in the filament region or in the cluster region when stacking galaxy sets of randomly re-positioned pairs. We therefore state that the treatment of sky coverage geometry in our algorithm is proper, and the filamentary as well as cluster overdensity observed in Section~\ref{section:maps} are real attributes of cluster pair fields.

\section{Results}
\label{section:results}

\subsection{Color Distributions}
  \label{section:color}
\begin{figure*}
\begin{center}

  \plotone{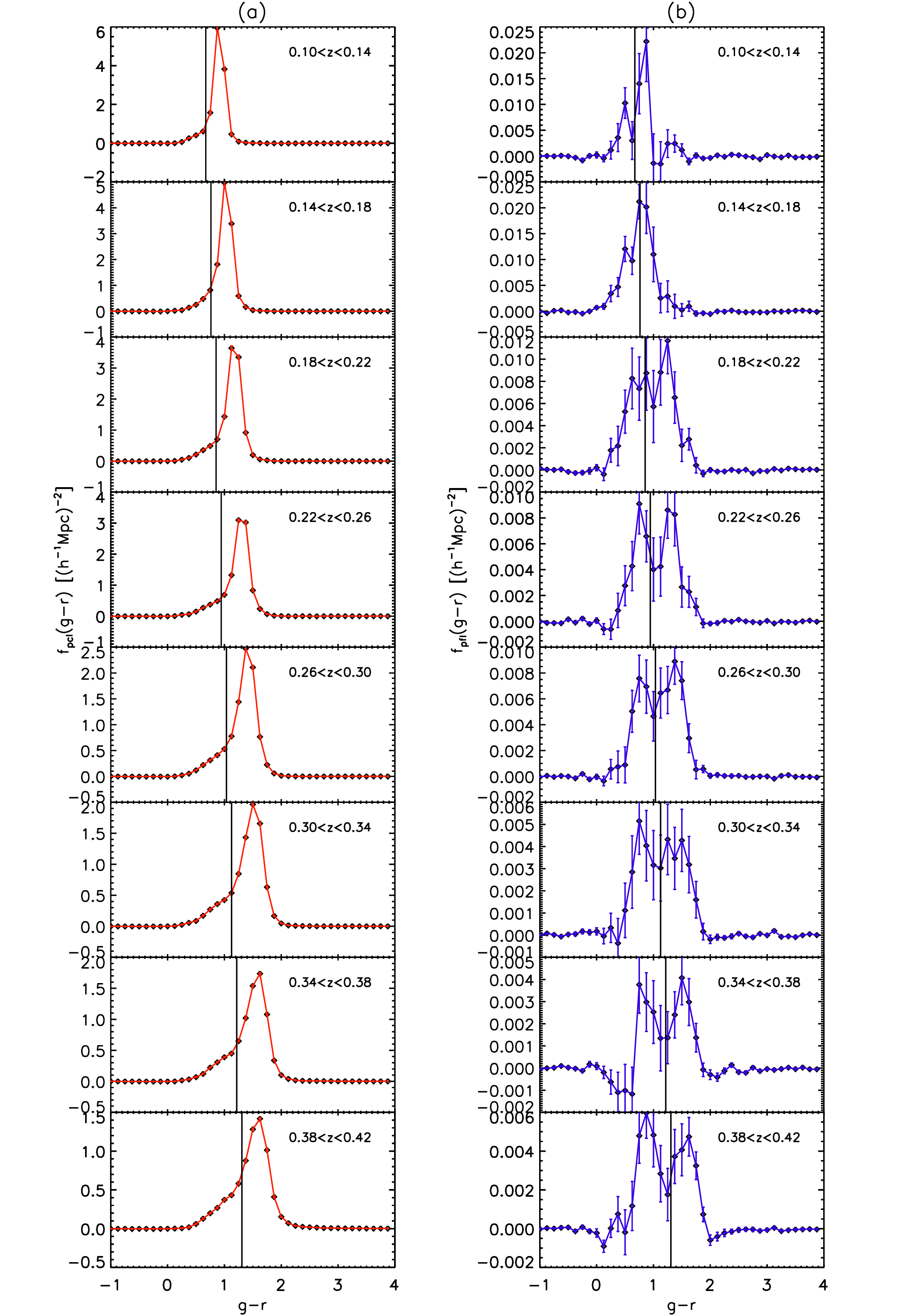}
  \caption{$g-r$ distributions of clusters (a) and filaments (b) in 8
    redshift bins. Filaments are bimodal with their color distribution
    and contain a larger blue galaxy bump than clusters. Also, the
    filament color distribution shows evidence of redshift evolution
    in the form of a blue galaxy population that becomes increasingly
    important at higher redshift. The vertical solid lines in these
    plots mark the blue/red galaxy color cut used in
    Section~\ref{section:fraction}. Note that the galaxy population
    sampled here is somewhat incomplete at $0.38<z<0.42$ (See
    discussion in Section~\ref{section:brightness_cut}). See the
    electronic edition of the journal for a color version of this
    figure.}
  \label{fig:gr100}
\end{center}
\end{figure*}

This section studies the $g-r$ color evolution of filaments at
$0.10<z<0.42$. For SDSS filters, despite the 4000\AA\ break transition
from g-band to r-band around $z \sim 0.38$, practically, $g-r$ works
almost equally well for separating blue late-type galaxies from red
early-type galaxies at $0.3<z<0.42$. \citet{2010ApJS..191..254H} have
a detailed discussion on using $g-r$ for detecting the red sequence in
cluster galaxies out to $z=0.42$.

With Equations~\ref{eq:fil}--\ref{eq:cl}, we compute the color
distributions of galaxy filaments, $f_\mathrm{pfl}(g-r)$, as well as
the color distributions of galaxy clusters, $f_\mathrm{pcl}(g-r)$, at
different redshifts. Results are shown in Figure~\ref{fig:gr100}. The
red sequence population is observed in both clusters and
filaments. Comparing to galaxy clusters, galaxy filaments have a
bimodal color distribution and a larger blue galaxy population. 

More interestingly, galaxy clusters and filaments both show evidence
of redshift evolution, with a blue cloud galaxy population increasing
steadily from $z=0.1$ to $z=0.4$. In galaxy clusters, such evolution
is observed as the Butcher-Oemler effect \citep{1978ApJ...219...18B,
  1978ApJ...226..559B, 1984ApJ...285..426B}, although the exact nature
of it is much debated. Many embrace this effect as testimony of
hierarchical clustering \citep{1995MNRAS.274..153K,
  1996MNRAS.283.1361B, 2001ApJ...547..609E, 2009MNRAS.400..937M,
  2009ApJ...698...83L}, but there are also strong voices that remain
critical \citep{1984ARA&A..22..185D,1998MNRAS.293..124S,
  1999ApJ...516..647A, 2004MNRAS.349..889A, 2006MNRAS.365..915A},
saying that either the error of measurements are underestimated, or
this effect is no more than selection bias. Nevertheless, our data
show a very strong redshift evolution of the cluster blue fraction and
suggest that a similar effect {\bf may }also exist in filaments. At
redshifts $z<0.2$, the blue galaxy cloud is only a small bump comparing
to the red sequence population. Moving to higher redshifts, the blue
cloud becomes more prominent, taking up the more than half of the
population at $z>0.28$.

\subsubsection{Blue Galaxy Fraction Evolution}
 \label{section:fraction}
\begin{figure*}
  \plotone{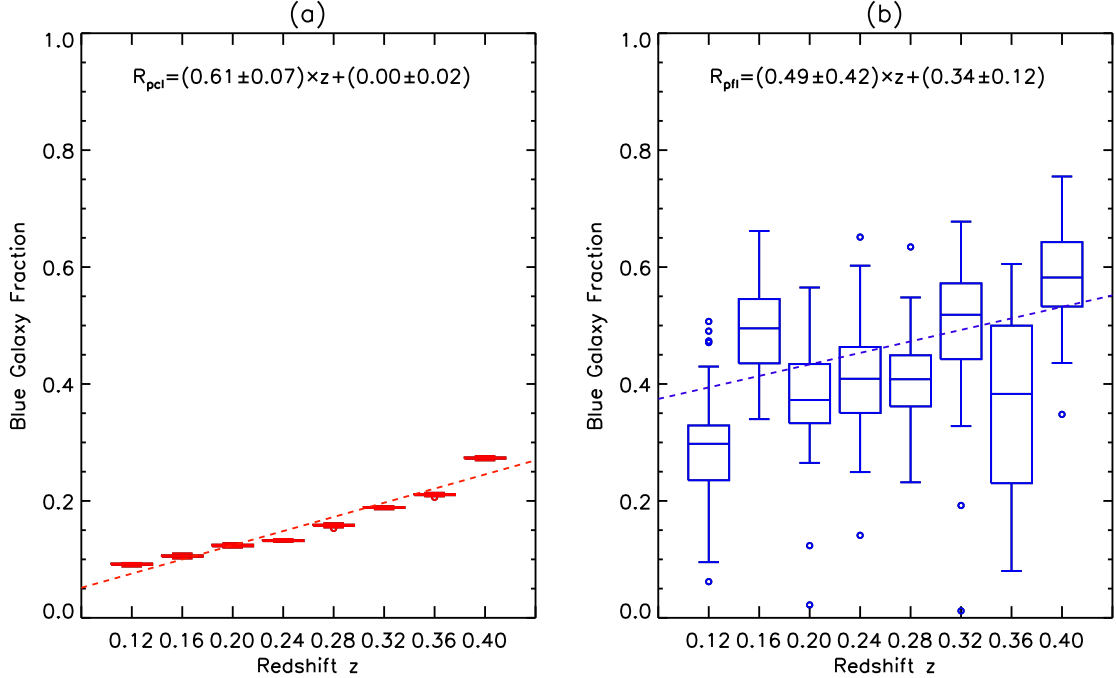}
  \caption{Redshift evolution of blue galaxy fraction in clusters (a)
    and filaments (b). These two figures show box plots of the blue
    galaxy fraction evaluation in clusters and filaments of 40
    bootstrapped stacks at each redshift bin. The bottom and top of
    the boxes represent the lower and upper quartiles, and the
    whiskers represent the minimum and maximum values of the data, or
    1.5 times the quartiles. Data points outside this range are
    plotted as open circles. We also fit the blue galaxy fraction
    linearly to the redshift (dashed lines). The blue galaxy
    population in clusters and filaments both seems to increase at higher
    redshift, but the change in filaments is less dramatic. Also note
    that the observed galaxy population is somewhat incomplete at
    $0.38<z<0.42$ (see discussion in
    Section~\ref{section:brightness_cut}). See the electronic edition
    of the journal for a color version of this figure.}
  \label{fig:fig_frac}
\end{figure*}

Figure~\ref{fig:gr100} shows that filaments and clusters both contain
blue and red galaxies. This figure suggests redshift evolution in the
filament galaxy population. This section quantifies this evolution by
measuring the blue galaxy fraction in filaments and clusters as a
function of redshift.

 To measure the blue galaxy fraction, we determine a blue/red galaxy
cut to separate the blue/red galaxy population. We take the minimum point between the blue and red galaxy peak in the bimodal filament color distribution, and fit it with a linear relation to redshift. In Figure~\ref{fig:gr100}, we have these linearly fitted values marked out as a vertical solid line in both cluster and filament color distributions, and use them as the blue/red
galaxy cut.   We tested other blue/red galaxy cut criteria, including using local minimum in the color distributions, double gaussian mixture fitting minimums and values extrapolated from passive evolution models, but the conclusions here do not change much. We compute the blue galaxy fraction in
filaments $R_\mathrm{pfl}$ and clusters $R_\mathrm{pcl}$ per redshift
bin by evaluating,
\begin{equation}
\label{eq:integration}
\begin{split}
  R_\mathrm{pfl} &= \frac{\int^{s_z}f_\mathrm{pfl}(g-r) 
    \mathrm{d}(g-r)}{\int f_\mathrm{pfl}(g-r) \mathrm{d}(g-r)}, \\
  R_\mathrm{pcl} &= \frac{\int^{s_z}f_\mathrm{pcl}(g-r) 
    \mathrm{d}(g-r)}{\int f_\mathrm{pcl}(g-r)  \mathrm{d}(g-r)}.
\end{split}
\end{equation}
Here $s_z$ is the $g-r$ color of the separation between blue and red
galaxies in this redshift bin.  The integrations in the above equations
are performed with a five-point Newton-Cotes integration formula on
the binned color distributions.

In Figure~\ref{fig:fig_frac} we show this blue galaxy fraction
estimation in filaments and clusters at different redshifts. We also
fit a linear evolution of the blue galaxy fraction with redshift (dashed lines)
using the MPFITEXY routine \citep{2010MNRAS.409.1330W} (dependent
on the MPFIT package from \citet{2009ASPC..411..251M}).  Our measurement
suggest that the blue galaxy fraction increases with redshift with a
linear fitting slope of $0.61\pm0.07$.

For filaments, the linear fit of the blue galaxy fraction as a
function of redshift slightly prefers a positive slope, $0.49\pm0.42$,
with a significantly higher intersect, which reflects the overall
bluer filament population.  We change our FL and C1 region definition to avoid including the cluster pre-processing area ($>4.2h^{-1}\mathrm{Mpc}$ away from cluster center), but do not find noticeable change in this effect. This indicates the independence of this redshift evolution feature on  pre-processing around clusters. We briefly discuss the implications of this
result in Section~\ref{sec:color-evolution}.

\begin{table*}
\begin{center}
\caption{Fitted Luminosity Function Parameters \label{tbl:lf}}
\begin{tabular}{cccccccc}
\tableline\tableline
& $z$ & $M_r$ cut/mag&$\phi_\mathrm{f}^*/(h^{3}\mathrm{Mpc}^{-3})$ & 
$M_\mathrm{r,f}^*-5\log_{10}h$/mag &
$\phi_\mathrm{c}^*/(h^{3}\mathrm{Mpc}^{-3})$ &
 $M_\mathrm{r, c}^*-5\log_{10}h$/mag\\
\tableline
&[0.10, 0.14] &$-24.2\le M_r \le-18.0$&$0.37 \pm 0.24$& $-23.49 \pm 5.40$& $5.18 \pm 0.13$ &$-21.544 \pm 0.033$\\
&[0.14, 0.18] &$-24.2\le M_r \le-18.6$&$0.32 \pm 0.18$& $-24.21 \pm 5.16$& $5.36 \pm 0.11$ &$-21.422 \pm 0.018$\\
&[0.18, 0.22] &$-24.2\le M_r \le-19.2$&$0.78 \pm 0.15$& $-20.87 \pm 0.25$& $5.21 \pm 0.15$ &$-21.369 \pm 0.018$\\
&[0.22, 0.26] &$-24.2\le M_r \le-19.6$&$0.73 \pm 0.17$& $-20.80 \pm 0.24$& $5.01 \pm 0.07$ &$-21.317 \pm 0.015$\\
&[0.26, 0.30] &$-24.2\le M_r \le-20.0$&$0.65 \pm 0.12$& $-20.76 \pm 0.20$& $4.86 \pm 0.06$ &$-21.229 \pm 0.014$\\
&[0.30, 0.34] &$-24.2\le M_r \le-20.4$&$0.41 \pm 0.11$& $-21.09 \pm 0.39$& $4.66 \pm 0.05$ &$-21.193 \pm 0.011$\\
&[0.34, 0.38] &$-24.2\le M_r \le-20.6$&$0.47 \pm 0.14$& $-20.66 \pm 0.31$& $4.67 \pm 0.06$ &$-21.137 \pm 0.012$\\
&[0.38, 0.42] &$-24.2\le M_r \le-21.0$&$0.45 \pm 0.21$& $-20.71 \pm 0.38$& $4.78 \pm 0.07$ &$-21.103 \pm 0.016$\\
\tableline
\end{tabular}
\end{center}

This table lists the fitted luminosity function parameters as
discussed in Section~\ref{section:luminosity}. The first column gives
the redshift range of each redshift bin, and the second column gives
the magnitude cut at this redshift, i.e., the luminosity function
fitting range. The errors of the parameters are evaluated through
fitting 40 resamplings of the galaxy sets being stacked at each
redshift bin. Error estimation of clusters is enlarged by a factor of
$\sqrt{5.7}$ since clusters are on-average oversampled by this
factor in the stack.
\end{table*}

\subsection{Luminosity Function}
 \label{section:luminosity}

\begin{figure*}
\begin{center}
\includegraphics[width=1.0\textwidth]{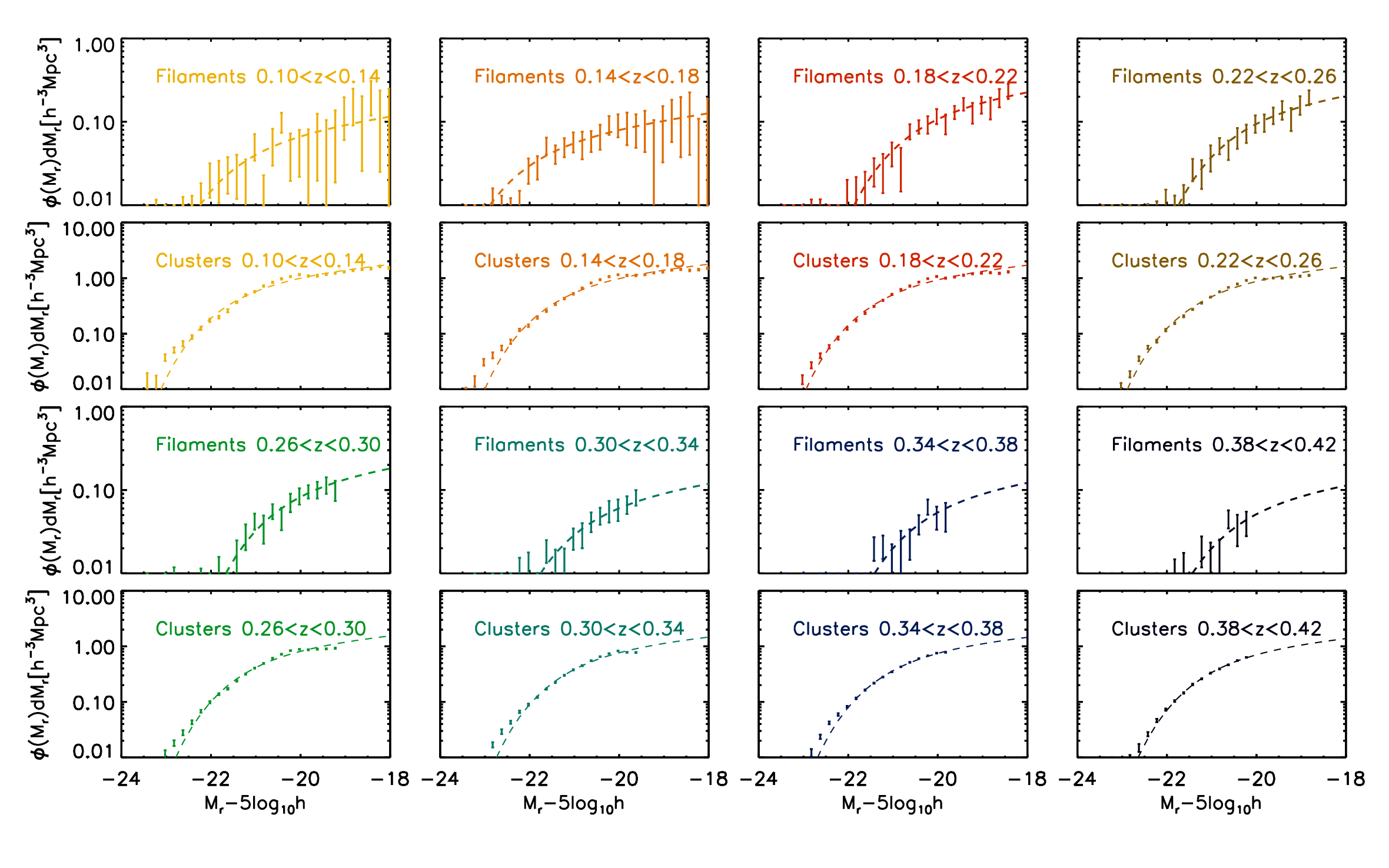}
\caption{Luminosity Distributions of filaments (1st and 3rd rows) and
  clusters (2nd and 4th rows) and fits to the Schechter function
  (dashed lines) using the MPFIT package
  \citep{2009ASPC..411..251M}. The fitting parameters are listed in
  Table~\ref{tbl:lf}. We do not fit for the $M_{r}-5\log_{10}h<-22.3$
  bright end of the luminosity distributions since a LRG population
  component is possibly present, and it luminosity distribution tend
  to deviate from the Schechter function
  \citep{2008ApJ...676..248Y}. See the electronic edition of the
  journal for a color version of this figure.}
  \label{fig:gmag}
\end{center}
\end{figure*}

In this section, we study the luminosity functions of the filament
and the cluster galaxy populations. Because our data does not allow us
to constrain the 3-dimensional distribution of cluster and in
particular of filament galaxies, we make the
following assumptions when computing the number density of galaxies:

\begin{enumerate}
\item $N$-body simulations predict that the density profiles fall off
  quickly between $1h^{-1}\mathrm{Mpc}$ and $2h^{-1}\mathrm{Mpc}$ from
  the intercluster axis \citep{2005MNRAS.359..272C}. Thus, a single
  filament will typically not occupy the whole FL region. We assume
  that one filament occupies 1/3 of the filament region defined in
  Section~\ref{section:maps}, and extends $\pm 1h^{-1}\mathrm{Mpc}$
  along the line of sight.
\item Since most galaxy clusters are at the scale of $1\,h^{-1}$\,Mpc,
  we assume that one cluster lies within the aperture of
  $\sqrt{x^2+y^2}<0.7\,h^{-1}$\,Mpc and also extends $\pm
  1\,h^{-1}$\,Mpc along the line of sight. Note that this does not
  imply a strong triaxiality of the cluster but rather limits the
  choice of cluster galaxies to those found in a cylinder fully
  contained within the cluster.
\item We assume that the purity of the gmBCG cluster catalog is
  $80\%$, which is the lower limit estimated for the gmBCG catalog
  \citep{2010ApJS..191..254H}.
\item We also assume that on average $20\%$ of the cluster pairs in
  our sample are connected by filaments. This again is a lower limit,
  estimated from the results of \citet{2005MNRAS.359..272C}.
\end{enumerate}
The assumptions listed here are clearly only rough approximations
adopted for convenience. We therefore stress that they only impact the
normalization of the luminosity function and not the determination of
the characteristic magnitudes as a function of
environment. Consequently, we caution the reader against comparing the
measured 3-d galaxy number density of filaments and clusters in this
paper to measurements obtained in other ways without taking these
approximations into account.

Under the above assumptions, the luminosity distributions of filaments
and clusters in one redshift bin, denoted by $\phi_\mathrm{pfl}(M_r)$
and $\phi_\mathrm{pcl}(M_r)$, respectively, are computed from $g_z(x,
y, M_r, g-r)$ and $cl_z( M_r, g-r)$ with the following equations:
\begin{equation}
\label{eq:lf_comp}
\begin{split}
\Phi_\mathrm{FL}(M_r) &= \frac{\sum\limits_{(i,j)\; \mathrm{in} \;
    \mathrm{FL} \;} \sum\limits_{l} g_z(X_i, Y_j, M_r,
  (g-r)_l)}{\sum\limits_{(i,j)\; \mathrm{in} \; 
    \mathrm{FL} \;}0.49(h^{-1}\mathrm{Mpc})^2}, \\
\Phi_\mathrm{CL} (M_r) &= \frac{\sum\limits_{l}
  cl_z( M_r, (g-r)_l)}{0.49(h^{-1}\mathrm{Mpc})^2},\\
\Phi_\mathrm{C1}(M_r) &= \frac{\sum\limits_{(i,j)\; \mathrm{in} \; 
    \mathrm{C1} \;}\sum\limits_{l} g_z(X_i, Y_j, M_r, 
  (g-r)_l)}{\sum\limits_{(i,j)\; \mathrm{in} \; \mathrm{C1} \;}0.49(h^{-1}\mathrm{Mpc})^2},\\
\Phi_\mathrm{C2}(M_r) &= \frac{\sum\limits_{(i,j)\; \mathrm{in} \; 
    \mathrm{C2} \;}\sum\limits_{l} g_z(X_i, Y_j, M_r,
  (g-r)_l)}{\sum\limits_{(i,j)\; 
    \mathrm{in} \; \mathrm{C2} \;}0.49(h^{-1}\mathrm{Mpc})^2},\\
\phi_\mathrm{pfl}(M_r)  &= \frac{\Phi_\mathrm{FL}(M_r) - 
  \Phi_\mathrm{C1}(M_r)}{\frac{1}{3}\times\frac{1}{20\%}\times 
  2\,h^{-1}\mathrm{Mpc}}, \\
\phi_\mathrm{pcl}(M_r) &= \frac{\Phi_\mathrm{CL}(M_r) -
  \Phi_\mathrm{C2}(M_r)}{\frac{1}{80\%}\times 2\,h^{-1}\mathrm{Mpc}}.
\end{split}
\end{equation}

In Figure~\ref{fig:gmag}, we plot the luminosity distributions of
filaments and clusters. We also fit the luminosity distribution of
filaments and clusters with a \citet{1976ApJ...203..297S} function:
\begin{equation}
\label{eq:fil_sc}
\phi(M_r)\mathrm{d}M_r = 0.4
\ln(10)\phi^{*}(10^{-0.4(M_r-M_{r}^*)})^{\alpha+1} 
e^{10^{-0.4(M_r-M_{r}^*)}}\mathrm{d}M_r.
\end{equation}
Because our data can only put weak constraints on the faint end of the
luminosity function, we chose to fix $\alpha = -1.2$, the value found
by \cite{2001AJ....121.2358B} and close to the measurements of
\cite{2005ApJ...620..618H} and \cite{2005A&A...433..415P}. The very
bright end of the luminosity distribution tends to depart from a
Schechter function \citep{2008ApJ...676..248Y}. We therefore restrict
our fits to galaxies with $M_{r} - 5\log_{10}h \ge -22.3$. We list the
results of our fits in Table~\ref{tbl:lf}. The normalizations,
$\Phi^*$ and r-band characteristic magnitude, $M_r^*$, of filaments
and clusters are denoted by $\phi^*_\mathrm{f}$, $M^*_\mathrm{r,f}$
and $\phi^*_\mathrm{c}$, $M^*_\mathrm{r,c}$, respectively.

In two individual redshift bins at $0.10<z<0.14$ and $0.14<z<0.18$ our
measurement of $M_\mathrm{f}^{*}$ has large errors. Visual inspection
of these fits suggests that the fixed faint end slope may not be a good
representation of the data in these bins. Especially in the low
redshift bins the data, however, is too noisy to leave the faint end
slope as a free parameter. We choose to ignore these two bins in our
subsequent discussion and analysis.

\subsection{Cluster Richness Dependence}
\label{section:richness}

This section studies the dependence of filament overdensity on cluster
richness. When stacking galaxy sets of clusters in the cluster pair
catalog, in addition to two redshift bins $0.10<z<0.26$ and
$0.26<z<0.42$, we further bin the cluster pairs into two richness
bins: one bin with both clusters in the pair having richness
estimation between 8 and 10 and one bin with both clusters in the pair
having richness larger than 10. Here, by richness, we mean the
$GM\_SCALED\_NGALS$ or $GM\_NGALS\_WEIGHTED$ richness estimation of
the gmBCG cluster catalog, whichever is recommended for a specific
galaxy cluster \citep{2010ApJS..191..254H}.

We evaluate the filament galaxy overdensity of different bins with a
quantity $\Sigma_\mathrm{pfl}$, which is computed with equations:
\begin{equation}
\label{eq:count}
\begin{split}
\Sigma_\mathrm{FL} &=\frac{\sum\limits_{(i,j)\; \mathrm{in} \; \mathrm{FL} \;}\sum\limits_{k, l} g_z(X_i, Y_j, (M_r)_k, (g-r)_l)}{\sum\limits_{(i,j)\; \mathrm{in} \; \mathrm{FL}}0.49(h^{-1}\mathrm{Mpc})^2}, \\
\Sigma_\mathrm{C1}&=\frac{\sum\limits_{(i,j)\; \mathrm{in} \; \mathrm{C1} \;}\sum\limits_{k, l} g_z(X_i, Y_j, (M_r)_k, (g-r)_l)}{\sum\limits_{(i,j)\; \mathrm{in} \; \mathrm{C1}}0.49(h^{-1}\mathrm{Mpc})^2},\\
\Sigma_\mathrm{pfl} &=\Sigma_\mathrm{FL}-\Sigma_\mathrm{C1}. \\
\end{split}
\end{equation}
This quantity is the 2-d projected galaxy number count per
$[h^{-1}\mathrm{Mpc}]^2$ of our detected filament galaxy overdensity.
However, this quantity should not be misinterpreted as the typical
surface number density of filament galaxies. Not every cluster pair
being stacked has a filament connection between them, and single
filaments typically occupy only a small subset of the FL
region. Consequently, the real projected galaxy count of filaments per
unit area can be significantly larger than the values given by
$\Sigma_\mathrm{pfl}$. In Table~\ref{tbl:rich}, we list the
$\Sigma_\mathrm{pfl}$ between cluster pairs at the above richness and
redshift binning. The $\Sigma_\mathrm{pfl}$ between richer cluster
pairs tends to be larger than the $\Sigma_\mathrm{pfl}$ between poorer
cluster pairs. Assuming the geometrical distribution of filaments does
not change with cluster richness, this result suggests that richer
cluster pairs tend to be connected by higher density filaments.

\begin{table}
\begin{center}
\caption{$\Sigma_\mathrm{pfl}$ $[(h^{-1}\mathrm{Mpc})^{-2}]$ versus richness at two redshift ranges \label{tbl:rich}}
\begin{tabular}{ccccr}
\tableline\tableline
& Richness range &  $0.10\leq z <0.26$ & $0.26\leq z <0.42$ & \\
\tableline
&8$\leq$Richness $<10$ & $0.053\pm0.030$ & $0.012\pm0.020$ & \\
&Richness $\geq10$ & $0.075\pm0.016$ & $0.061\pm0.008$ & \\
\tableline
\end{tabular}
\end{center}
\end{table}

\subsection{Spatial Distribution}
\label{section:spatial}
\begin{table*}
  \begin{center}
    \caption{Definitions of the FL and C1 boxes used in Section\label{tbl:spatial}}
    \begin{tabular}{cccc}
      \tableline\tableline
      & Number & FL & C1 \\
      \tableline
      &1 & $2.1h^{-1}\mathrm{Mpc}<x<10.5h^{-1}\mathrm{Mpc}$, $0^{\circ}\le|\theta=\arctan(y/x)|< 15^{\circ}$
      & $-10.5h^{-1}\mathrm{Mpc}<x<-2.1h^{-1}\mathrm{Mpc}$,  $0^{\circ}\le|\theta=\arctan(y/x)|< 15^{\circ}$ \\
      &2 & $2.1h^{-1}\mathrm{Mpc}<x<10.5h^{-1}\mathrm{Mpc}$, $0^{\circ}\le|\theta=\arctan(y/x)|< 30^{\circ}$
      & $-10.5h^{-1}\mathrm{Mpc}<x<-2.1h^{-1}\mathrm{Mpc}$, $0^{\circ}\le|\theta=\arctan(y/x)|< 30^{\circ}$ \\
      &3 & $2.1h^{-1}\mathrm{Mpc}<x<10.5h^{-1}\mathrm{Mpc}$, $15^{\circ}\le|\theta=\arctan(y/x)|< 45^{\circ}$
      & $-10.5h^{-1}\mathrm{Mpc}<x<-2.1h^{-1}\mathrm{Mpc}$,  $15^{\circ}\le|\theta=\arctan(y/x)|< 45^{\circ}$ \\
      &4 & $2.1h^{-1}\mathrm{Mpc}<x<10.5h^{-1}\mathrm{Mpc}$, $30^{\circ}\le|\theta=\arctan(y/x)|< 60^{\circ}$
      & $-10.5h^{-1}\mathrm{Mpc}<x<-2.1h^{-1}\mathrm{Mpc}$, $30^{\circ}\le|\theta=\arctan(y/x)|< 60^{\circ}$ \\
      \tableline
    \end{tabular}
  \end{center}
\end{table*}

\begin{figure*}
  \begin{center}
    \plotone{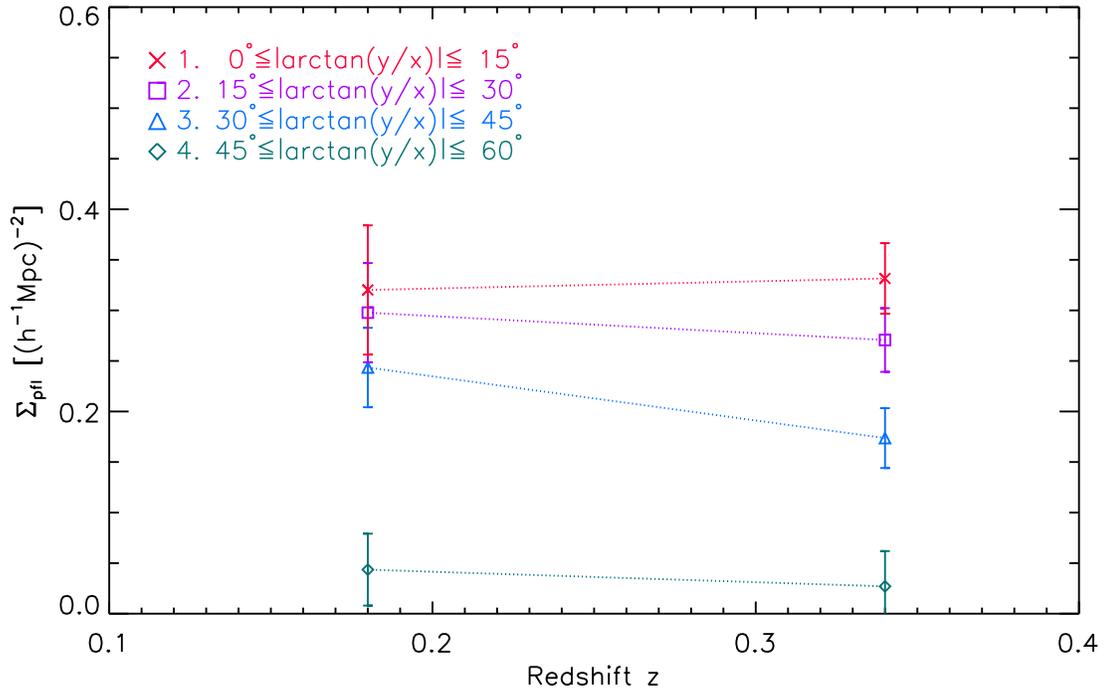}
    \caption{The 2-d projected galaxy number density of filaments,
      $\Sigma_\mathrm{pfl}$, with FL and C1 boxes placed at different
      extension angles. The exact definition of FL and C1 are listed
      in Table~\ref{tbl:spatial}. Filament galaxies still can be seen
      at increasing angles off the intercluster pair axis, but
      eventually die out when the extension angle from this axis is
      larger than $45^{\circ}$. See the electronic edition of the
      journal for a color version of this figure.}
    \label{fig:angle}
  \end{center}
\end{figure*}

Galaxy filaments can be warped and do not always strictly lie along
the intercluster axis
\citep{2004MNRAS.354L..61P,2005MNRAS.359..272C}. Therefore,
significant filamentary overdensities still can be seen off the
intercluster axis. In this section, we rotate the FL region and the C1
region away from the intercluster axis as described in
Table~\ref{tbl:spatial} and compare the surface number density of
galaxies $\Sigma_\mathrm{pfl}$ at two redshift ranges $[0.10,0.26]$ and
$[0.26,0.42]$. Our result is shown in Figure~\ref{fig:angle}. As we
move away from the intercluster axis, the observed filament population
gradually falls off. At $\theta>45^{\circ}$, the filament galaxy
overdensity is almost consistent with zero. 

\section{Summary and Discussion}
\label{section:overview}

In this paper we developed an algorithm to detect and study the
properties of galaxies in filaments
statistically. Our method makes use of the fact that close pairs of
galaxy clusters have a large probability of being connected by
filaments. This is a robust prediction of numerical simulations of
structure formation \citep[e.g.,][]{2005MNRAS.359..272C} and has been
confirmed observationally in galaxy redshift surveys
\citep[e.g.,][]{2004MNRAS.354L..61P}. This allows us to study filament
galaxies statistically without the need to spectroscopically identify
them. Our algorithm is applicable to all photometric surveys, in which
large numbers of galaxy clusters can be found. As a result we can
study the properties of representative filament galaxies out to much
higher redshifts than is currently possible with wide-field
spectroscopic surveys.

We applied our new method to the gmBCG galaxy cluster catalog and the
photometric SDSS DR8 BOSS galaxy catalog. Although the projected
filament galaxy densities exceed that of the surrounding field by only
$\sim 2\%$, we can detect these relative overdensities at high
significance (often $\sim 5\sigma$) in redshift bins of thickness
$\Delta z = 0.04$ at $0.10 < z < 0.42$. The limit of this redshift
range is governed by the gmBCG cluster finding algorithm and the small
survey volume in the local Universe at the low redshift end, and by
the flux limit of SDSS at the high redshift end. Future deep
wide-field surveys like the Dark Energy
Survey\footnote{\texttt{http://www.darkenergeysurvey.org/}} will allow
us to push this method to higher redshifts.

As an application of our method, we study the color and luminosity
distributions of galaxy filaments, as well as the dependence of the
filament richness on the richness of the clusters to which they are
connected. Finally, we put limits on the warping angles of large-scale
structure filaments.

As verification additional to that already presented in
Section~\ref{section:methods}, we also apply our algorithm to the
Millennium Simulation \citep{2005Natur.435..629S}. We present this in
the appendix and only mention here that such tests support the
efficacy of our methods and also confirm previous suggestions that the
Millennium Simulation is overpopulated with red
galaxies \citep{2011MNRAS.413..101G}.

\subsection{Color Evolution}
\label{sec:color-evolution}

We find that filament galaxies have a clear bimodal color
distribution. At redshift $0.10<z<0.14$, the filament $g-r$ color
distribution is qualitatively comparable to the $u-r$ color
distribution of bright wall galaxies at $z<0.107$ presented in
\citet{2012MNRAS.426.3041H}: both being bimodal, and the blue cloud
galaxy taking up $\sim 1/3$ of the whole population.

 We find clear indications for a Butcher-Oemler effect in galaxy
clusters at very high significance. Fitting a linear function to the
blue galaxy fraction we find that it increases with redshift with a
slope of $0.61\pm0.07$ over the redshift range studied. This is
comparable to the results of \citet{2009ApJ...699.1333H}, which showed that from $z=0.28$ to $z=0.2$,
there is a $\sim 5\%$ decrease in the blue galaxy fraction, which our linear fit agrees with. Care needs to be exercised when comparing our measurement to many other works studying the Butcher-Oemler effect, though. For example, \citeauthor{2001ApJ...548L.143M} studied
the Butcher Oemler effect with 295 Abell clusters, and showed that the blue fraction in these clusters increases with a slope of $1.24\pm0.07$ at redshift range $z<0.4$. Since \citeauthor{2001ApJ...548L.143M} adopted different cluster aperture ($0.49$\,$h^{-1}\mathrm{Mpc}$ instead of $0.7$\,$h^{-1}\mathrm{Mpc}$), 
brighter magnitude limit ($M^*_r-1<M_r<M^*_r+1$, which is approximately $-22.2<M_r<-19.2$
according to our measurements of $M_r^*$) and completely different blue/red galaxy cut (rest frame $g-r$ color 0.2 magnitude fainter than the red sequence ridgeline), it is not particularly
surprising that we have different estimation on the blue galaxy fraction evolution strength, but a more specified study on this effect in clusters may
wish to refine the choice of cluster aperture, brightness limit, and blue/red galaxy cut used here.

We marginally detect an evolution in the blue fraction of filament
galaxies with a linear slope of $0.49\pm 0.42$. This indicates that there might be a trend that
the population of blue galaxies in filaments is increasing at higher
redshift, but this redshift evolution feature is weaker than in
clusters. The implications of such an observation are complicated,
though. Should galaxy filaments be considered as ensemble of galaxy
groups and galaxies in less dense environments, our observation can be the result of a Butcher-Oemler effect in galaxy groups, as discussed
in \cite{2012ApJ...749..150L}. \citeauthor{2012ApJ...749..150L}
measured the red/blue galaxy fraction in galaxy groups at different mass
ranges and found that galaxy groups of all different masses show a
trend of increasing blue galaxy fraction at higher redshifts. Their
measurements of the blue galaxy fraction in galaxy groups at their
lowest total steller mass bin
$\log(M_{*,\mathrm{grp}}/M_{\odot})<11.2$ are $\sim 48\%$ at $z=0.25$
and $\sim 52\%$ at $z=0.35$, is consistent with
our interpolated values $\sim 46.3\%$ at $z=0.25$ and $\sim 51.2\%$ at
$z=0.35$.   The possible redshift
  evolution of the blue galaxy fraction in filaments can also be a ``by-product'' of changes in filament galaxy contents or
  morphologies at different redshifts.  Our results on
the evolution of the blue galaxy fraction in filaments are currently
limited by the signal-to-noise ratio per redshift bin. Future deep
surveys will give us a longer lever arm in redshift to study this
effect in more detail. We also expect that improvements in galaxy
cluster selection will lead to an enhanced catalog purity at all
redshifts. This will improve the SNR at all redshifts.

\subsection{Luminosity Function}
\label{sec:luminosity-function}

We measured the characteristic r-band magnitude of cluster and
filament galaxies in redshift bins by fitting a standard
\citeauthor{1976ApJ...203..297S} luminosity function with a fixed
faint end slope of $\alpha= -1.2$. For cluster galaxies,
\citet{2005A&A...433..415P}, who studied the RASS-SDSS clusters to
$z=0.26$, is a suitable comparison to our work. Their measurements of
the cluster r-band characteristic magnitude at $z<0.26$ are $-21.35\pm
0.19$\,mag or $-21.40\pm 0.20$\,mag, depending on the choice of
background. This is in good agreement with our measurements. An
increase of $M^*_\mathrm{r}$ of cluster galaxies with redshift is seen
at high statistical significance. This result, however, may be
affected by a potential redshift dependence of the purity of the gmBCG
catalog, as well as a redshift dependent richness threshold in the
cluster finder. A detailed study of these factors is beyond the scope
of the present work and the redshift evolution of $M^*$ should be
treated with caution.

In redshift bins where we obtain reliable measurements, the r-band
$M^*$ of filament galaxies is significantly lower than that of
cluster galaxies. We find no significant redshift evolution of
$M^*_\mathrm{r,f}$.

\begin{table}
\begin{center}
\caption{SDSS r-band characteristic magnitude measurements \label{tbl:mr}}
\begin{tabular}{cccc}
\tableline\tableline
&   & $M^*_r$ Measurements\\
\tableline
&Filament $M_r^*$ (our measurement) & $-20.85\pm0.11$\\
&\cite{2001AJ....121.2358B} $M_r^*$ & $-20.83\pm0.03$\\
&\cite{2009MNRAS.399.1106M} $^{0.1}M_r^*$ & $-20.71\pm0.04$ \tablenotemark{a}\\
\tableline
\end{tabular}
\end{center}
$^a$ Note that this magnitude is measured for the SDSS r-band blue
shifted by 0.1, i.e., $^{0.1}r$. \citep{2003AJ....125.2348B,
  2003ApJ...592..819B}, and are not directly comparable to the rest in
the table. At $z=0.1$, the K-correction to such bands are
$-2.5\log_{10}1.1=-0.1035$, but the K-correction for our system at
$z=0.1$ ranges from $-0.06$ to $0.37$. Here, we make a simple
assumption that there is a uniform offset between the galaxy absolute
magnitude measured in the $^{0.1}r$ band and the regular $r$
band. Since the  the K-correction we use at $z=0.1$ scatters around
0.08, which is the K-correction for galaxies of $z=0.1$ and $g-r=0.8$
using the \citet{2010MNRAS.405.1409C} approximation, we assume the
offset between the two systems to be $0.18$. The r-band characteristic
magnitude measured  in \cite{2009MNRAS.399.1106M} is now closer to our
filament measurement. 
\end{table}

When we stack the luminosity distributions of filament galaxies over
$0.18<z<0.42$, we obtain a characteristic luminosity of filament
galaxies of $M_\mathrm{r,f}^{*} = -20.85\pm0.11$\,mag. This value is
compatible with two r-band $M^{*}$ measurements (listed in
Table~\ref{tbl:mr}) of SDSS galaxies, in which galaxies were not
separated by their environment. This re-enforces the result of
\citet{2010MNRAS.408.2163A} that most matter and thus most galaxies
live in filaments and not in clusters. Consequently, the filament
environment is expected to dominate globally averaged
measurements. Our value is also close to the r-band $M^{*}$
measurement of wall galaxies in \citet{2005ApJ...620..618H}, who found
$M^{*}=-20.62\pm0.08$\,mag. The filament galaxies in this work fall
into the category of wall galaxies in \citet{2005ApJ...620..618H}
(galaxies in denser environments than extremely underdense cosmic
voids). Not surprisingly, filament galaxies are significantly brighter
than void galaxies, whose r-band $M^{*}$ was measured to be
$-19.74\pm0.11$\,mag by \citet{2005ApJ...620..618H}.

\subsection{Filament Richness and Geometry}
\label{sec:filam-richn-geom}
We find that richer clusters are connected to richer filaments. This
is consistent with description of structure formation given by
\citet{1996Natur.380..603B}. In this picture, superclusters with
pronounced filaments between their massive constituent clusters are
the result of rare initial density peaks. In turn, less massive
clusters were ``disadvantaged'' from the beginning by being more
isolated and having fewer and less dense filaments connected to
them. Because we select up to 5 neighbors for any cluster to form
cluster pairs with, independent of the cluster richness, both factors
-- the density and the number of filaments -- affects the measured
surface density of galaxies. Consistent with this expectation, we
find that the correlation of filament richness with cluster richness
exists at two redshift ranges studied here.

Finally, we also studied the distribution of matter in what we call
the filament region between the clusters. Slightly more than half of
all filaments are curved and some connect the clusters at which they
terminate with offsets from the intercluster axis. These reasons make
it impossible to disentangle the density profile of filaments from the
distribution of filament geometries in our stacks. Nevertheless, the
galaxy density in the filament region should decrease as one moves
away from the intercluster axis. We verified this by varying the
position angle of the filament region when rotating it around the
cluster center. We find that filaments are unlikely to be warped by
more than $45^{\circ}$ from the intercluster axis.

\subsection{Outlook}
\label{sec:outlook}
This stacking approach to studying filaments is not only applicable
with optical galaxy data, but can also be applied to weak lensing
measurements, Sunyaev-Zeldovich effect observations, and X-ray
emission from the warm-hot intercluster medium, provided appropriately
large survey data exist. We look forward to its broader application in
the era of combined data from deep wide-field optical, SZ, and X-ray
surveys.

\acknowledgements The authors are pleased to acknowledge generous
support from NSF grant AST-0807304. JPD was also partially supported
by the German Science Foundation (DFG) through the Transregio 33 and
the Cluster of Excellence ``Origin and Structure of the Universe'',
funded by the Excellence Initiative of the Federal Government of
Germany, EXC project number 153. We are also very grateful to Jeeseon
Song, Heidi Wu, Adam Sypnieski and Risa Wechsler for many helpful
discussions. Funding for the SDSS and SDSS-II has been provided by the
Alfred P. Sloan Foundation, the Participating Institutions, the
National Science Foundation, the U.S. Department of Energy, the
National Aeronautics and Space Administration, the Japanese
Monbukagakusho, the Max Planck Society, and the Higher Education
Funding Council for England. The SDSS Web Site is
http://www.sdss.org/.

The SDSS is managed by the Astrophysical Research Consortium for the
Participating Institutions. The Participating Institutions are the
American Museum of Natural History, Astrophysical Institute Potsdam,
University of Basel, University of Cambridge, Case Western Reserve
University, University of Chicago, Drexel University, Fermilab, the
Institute for Advanced Study, the Japan Participation Group, Johns
Hopkins University, the Joint Institute for Nuclear Astrophysics, the
Kavli Institute for Particle Astrophysics and Cosmology, the Korean
Scientist Group, the Chinese Academy of Sciences (LAMOST), Los Alamos
National Laboratory, the Max-Planck-Institute for Astronomy (MPIA),
the Max-Planck-Institute for Astrophysics (MPA), New Mexico State
University, Ohio State University, University of Pittsburgh,
University of Portsmouth, Princeton University, the United States
Naval Observatory, and the University of Washington.

\acknowledgments
\bibliography{filament_stacking}

\begin{thebibliography}{66}
\expandafter\ifx\csname natexlab\endcsname\relax\def\natexlab#1{#1}\fi

\bibitem[{{Abazajian} {et~al.}(2009){Abazajian}, {Adelman-McCarthy},
  {Ag{\"u}eros}, {Allam}, {Allende Prieto}, {An}, {Anderson}, {Anderson},
  {Annis}, {Bahcall}, \& et~al.}]{2009ApJS..182..543A}
{Abazajian}, K.~N., {Adelman-McCarthy}, J.~K., {Ag{\"u}eros}, M.~A., {et~al.}
  2009, \apjs, 182, 543

\bibitem[{{Ahn} {et~al.}(2012){Ahn}, {Alexandroff}, {Allende Prieto},
  {Anderson}, {Anderton}, {Andrews}, {Aubourg}, {Bailey}, {Balbinot}, {Barnes},
  {Bautista}, {Beers}, {Beifiori}, {Berlind}, {Bhardwaj}, {Bizyaev}, {Blake},
  {Blanton}, {Blomqvist}, {Bochanski}, \& et~al.}]{2012ApJS..203...21A}
{Ahn}, C.~P., {Alexandroff}, R., {Allende Prieto}, C., {et~al.} 2012, \apjs,
  203, 21

\bibitem[{{Aihara} {et~al.}(2011){Aihara}, {Allende Prieto}, {An}, {Anderson},
  {Aubourg}, {Balbinot}, {Beers}, {Berlind}, {Bickerton}, {Bizyaev}, {Blanton},
  {Bochanski}, {Bolton}, {Bovy}, {Brandt}, {Brinkmann}, {Brown}, {Brownstein},
  {Busca}, {Campbell}, {Carr}, {Chen}, {Chiappini}, {Comparat}, {Connolly},
  {Cortes}, {Croft}, {Cuesta}, {da Costa}, {Davenport}, {Dawson}, {Dhital},
  {Ealet}, {Ebelke}, {Edmondson}, {Eisenstein}, {Escoffier}, {Esposito},
  {Evans}, {Fan}, {Femen{\'{\i}}a Castell{\'a}}, {Font-Ribera}, {Frinchaboy},
  {Ge}, {Gillespie}, {Gilmore}, {Gonz{\'a}lez Hern{\'a}ndez}, {Gott}, {Gould},
  {Grebel}, {Gunn}, {Hamilton}, {Harding}, {Harris}, {Hawley}, {Hearty}, {Ho},
  {Hogg}, {Holtzman}, {Honscheid}, {Inada}, {Ivans}, {Jiang}, {Johnson},
  {Jordan}, {Jordan}, {Kazin}, {Kirkby}, {Klaene}, {Knapp}, {Kneib},
  {Kochanek}, {Koesterke}, {Kollmeier}, {Kron}, {Lampeitl}, {Lang}, {Le Goff},
  {Lee}, {Lin}, {Long}, {Loomis}, {Lucatello}, {Lundgren}, {Lupton}, {Ma},
  {MacDonald}, {Mahadevan}, {Maia}, {Makler}, {Malanushenko}, {Malanushenko},
  {Mandelbaum}, {Maraston}, {Margala}, {Masters}, {McBride}, {McGehee},
  {McGreer}, {M{\'e}nard}, {Miralda-Escud{\'e}}, {Morrison}, {Mullally},
  {Muna}, {Munn}, {Murayama}, {Myers}, {Naugle}, {Neto}, {Nguyen}, {Nichol},
  {O'Connell}, {Ogando}, {Olmstead}, {Oravetz}, {Padmanabhan},
  {Palanque-Delabrouille}, {Pan}, {Pandey}, {P{\^a}ris}, {Percival},
  {Petitjean}, {Pfaffenberger}, {Pforr}, {Phleps}, {Pichon}, {Pieri}, {Prada},
  {Price-Whelan}, {Raddick}, {Ramos}, {Reyl{\'e}}, {Rich}, {Richards}, {Rix},
  {Robin}, {Rocha-Pinto}, {Rockosi}, {Roe}, {Rollinde}, {Ross}, {Ross},
  {Rossetto}, {S{\'a}nchez}, {Sayres}, {Schlegel}, {Schlesinger}, {Schmidt},
  {Schneider}, {Sheldon}, {Shu}, {Simmerer}, {Simmons}, {Sivarani}, {Snedden},
  {Sobeck}, {Steinmetz}, {Strauss}, {Szalay}, {Tanaka}, {Thakar}, {Thomas},
  {Tinker}, {Tofflemire}, {Tojeiro}, {Tremonti}, {Vandenberg}, {Vargas
  Maga{\~n}a}, {Verde}, {Vogt}, {Wake}, {Wang}, {Weaver}, {Weinberg}, {White},
  {White}, {Yanny}, {Yasuda}, {Yeche}, \& {Zehavi}}]{2011ApJS..193...29A}
{Aihara}, H., {Allende Prieto}, C., {An}, D., {et~al.} 2011, \apjs, 193, 29

\bibitem[{{Andreon} \& {Ettori}(1999)}]{1999ApJ...516..647A}
{Andreon}, S., \& {Ettori}, S. 1999, \apj, 516, 647

\bibitem[{{Andreon} {et~al.}(2004){Andreon}, {Lobo}, \&
  {Iovino}}]{2004MNRAS.349..889A}
{Andreon}, S., {Lobo}, C., \& {Iovino}, A. 2004, \mnras, 349, 889

\bibitem[{{Andreon} {et~al.}(2006){Andreon}, {Quintana}, {Tajer}, {Galaz}, \&
  {Surdej}}]{2006MNRAS.365..915A}
{Andreon}, S., {Quintana}, H., {Tajer}, M., {Galaz}, G., \& {Surdej}, J. 2006,
  \mnras, 365, 915

\bibitem[{{Arag{\'o}n-Calvo} {et~al.}(2010){Arag{\'o}n-Calvo}, {van de
  Weygaert}, \& {Jones}}]{2010MNRAS.408.2163A}
{Arag{\'o}n-Calvo}, M.~A., {van de Weygaert}, R., \& {Jones}, B.~J.~T. 2010,
  \mnras, 408, 2163

\bibitem[{{Baugh} {et~al.}(1996){Baugh}, {Cole}, \&
  {Frenk}}]{1996MNRAS.283.1361B}
{Baugh}, C.~M., {Cole}, S., \& {Frenk}, C.~S. 1996, \mnras, 283, 1361

\bibitem[{{Bertschinger} \& {Gelb}(1991)}]{1991ComPh...5..164B}
{Bertschinger}, E., \& {Gelb}, J.~M. 1991, Computers in Physics, 5, 164

\bibitem[{{Blaizot} {et~al.}(2005){Blaizot}, {Wadadekar}, {Guiderdoni},
  {Colombi}, {Bertin}, {Bouchet}, {Devriendt}, \&
  {Hatton}}]{2005MNRAS.360..159B}
{Blaizot}, J., {Wadadekar}, Y., {Guiderdoni}, B., {et~al.} 2005, \mnras, 360,
  159

\bibitem[{{Blanton} \& {Roweis}(2007)}]{2007AJ....133..734B}
{Blanton}, M.~R., \& {Roweis}, S. 2007, \aj, 133, 734

\bibitem[{{Blanton} {et~al.}(2001){Blanton}, {Dalcanton}, {Eisenstein},
  {Loveday}, {Strauss}, {SubbaRao}, {Weinberg}, {Anderson}, {Annis}, {Bahcall},
  {Bernardi}, {Brinkmann}, {Brunner}, {Burles}, {Carey}, {Castander},
  {Connolly}, {Csabai}, {Doi}, {Finkbeiner}, {Friedman}, {Frieman}, {Fukugita},
  {Gunn}, {Hennessy}, {Hindsley}, {Hogg}, {Ichikawa}, {Ivezi{\'c}}, {Kent},
  {Knapp}, {Lamb}, {Leger}, {Long}, {Lupton}, {McKay}, {Meiksin}, {Merelli},
  {Munn}, {Narayanan}, {Newcomb}, {Nichol}, {Okamura}, {Owen}, {Pier}, {Pope},
  {Postman}, {Quinn}, {Rockosi}, {Schlegel}, {Schneider}, {Shimasaku},
  {Siegmund}, {Smee}, {Snir}, {Stoughton}, {Stubbs}, {Szalay}, {Szokoly},
  {Thakar}, {Tremonti}, {Tucker}, {Uomoto}, {Vanden Berk}, {Vogeley},
  {Waddell}, {Yanny}, {Yasuda}, \& {York}}]{2001AJ....121.2358B}
{Blanton}, M.~R., {Dalcanton}, J., {Eisenstein}, D., {et~al.} 2001, \aj, 121,
  2358

\bibitem[{{Blanton} {et~al.}(2003{\natexlab{a}}){Blanton}, {Brinkmann},
  {Csabai}, {Doi}, {Eisenstein}, {Fukugita}, {Gunn}, {Hogg}, \&
  {Schlegel}}]{2003AJ....125.2348B}
{Blanton}, M.~R., {Brinkmann}, J., {Csabai}, I., {et~al.} 2003{\natexlab{a}},
  \aj, 125, 2348

\bibitem[{{Blanton} {et~al.}(2003{\natexlab{b}}){Blanton}, {Hogg}, {Bahcall},
  {Brinkmann}, {Britton}, {Connolly}, {Csabai}, {Fukugita}, {Loveday},
  {Meiksin}, {Munn}, {Nichol}, {Okamura}, {Quinn}, {Schneider}, {Shimasaku},
  {Strauss}, {Tegmark}, {Vogeley}, \& {Weinberg}}]{2003ApJ...592..819B}
{Blanton}, M.~R., {Hogg}, D.~W., {Bahcall}, N.~A., {et~al.} 2003{\natexlab{b}},
  \apj, 592, 819

\bibitem[{{Bond} {et~al.}(1996){Bond}, {Kofman}, \&
  {Pogosyan}}]{1996Natur.380..603B}
{Bond}, J.~R., {Kofman}, L., \& {Pogosyan}, D. 1996, \nat, 380, 603

\bibitem[{{Buote} {et~al.}(2009){Buote}, {Zappacosta}, {Fang}, {Humphrey},
  {Gastaldello}, \& {Tagliaferri}}]{2009ApJ...695.1351B}
{Buote}, D.~A., {Zappacosta}, L., {Fang}, T., {et~al.} 2009, \apj, 695, 1351

\bibitem[{{Butcher} \& {Oemler}(1978{\natexlab{a}})}]{1978ApJ...219...18B}
{Butcher}, H., \& {Oemler}, Jr., A. 1978{\natexlab{a}}, \apj, 219, 18

\bibitem[{{Butcher} \& {Oemler}(1978{\natexlab{b}})}]{1978ApJ...226..559B}
---. 1978{\natexlab{b}}, \apj, 226, 559

\bibitem[{{Butcher} \& {Oemler}(1984)}]{1984ApJ...285..426B}
---. 1984, \apj, 285, 426

\bibitem[{{Chilingarian} {et~al.}(2010){Chilingarian}, {Melchior}, \&
  {Zolotukhin}}]{2010MNRAS.405.1409C}
{Chilingarian}, I.~V., {Melchior}, A.-L., \& {Zolotukhin}, I.~Y. 2010, \mnras,
  405, 1409

\bibitem[{{Colberg} {et~al.}(2005){Colberg}, {Krughoff}, \&
  {Connolly}}]{2005MNRAS.359..272C}
{Colberg}, J.~M., {Krughoff}, K.~S., \& {Connolly}, A.~J. 2005, \mnras, 359,
  272

\bibitem[{{Colless} {et~al.}(2001){Colless}, {Dalton}, {Maddox}, {Sutherland},
  {Norberg}, {Cole}, {Bland-Hawthorn}, {Bridges}, {Cannon}, {Collins}, {Couch},
  {Cross}, {Deeley}, {De Propris}, {Driver}, {Efstathiou}, {Ellis}, {Frenk},
  {Glazebrook}, {Jackson}, {Lahav}, {Lewis}, {Lumsden}, {Madgwick}, {Peacock},
  {Peterson}, {Price}, {Seaborne}, \& {Taylor}}]{2001MNRAS.328.1039C}
{Colless}, M., {Dalton}, G., {Maddox}, S., {et~al.} 2001, \mnras, 328, 1039

\bibitem[{{Cunha} {et~al.}(2009){Cunha}, {Lima}, {Oyaizu}, {Frieman}, \&
  {Lin}}]{2009MNRAS.396.2379C}
{Cunha}, C.~E., {Lima}, M., {Oyaizu}, H., {Frieman}, J., \& {Lin}, H. 2009,
  \mnras, 396, 2379

\bibitem[{{Dav{\'e}} {et~al.}(2001){Dav{\'e}}, {Cen}, {Ostriker}, {Bryan},
  {Hernquist}, {Katz}, {Weinberg}, {Norman}, \& {O'Shea}}]{2001ApJ...552..473D}
{Dav{\'e}}, R., {Cen}, R., {Ostriker}, J.~P., {et~al.} 2001, \apj, 552, 473

\bibitem[{{Davis} {et~al.}(1985){Davis}, {Efstathiou}, {Frenk}, \&
  {White}}]{1985ApJ...292..371D}
{Davis}, M., {Efstathiou}, G., {Frenk}, C.~S., \& {White}, S.~D.~M. 1985, \apj,
  292, 371

\bibitem[{{Dietrich} {et~al.}(2012){Dietrich}, {Werner}, {Clowe}, {Finoguenov},
  {Kitching}, {Miller}, \& {Simionescu}}]{2012Natur.487..202D}
{Dietrich}, J.~P., {Werner}, N., {Clowe}, D., {et~al.} 2012, \nat, 487, 202

\bibitem[{{Dressler}(1984)}]{1984ARA&A..22..185D}
{Dressler}, A. 1984, \araa, 22, 185

\bibitem[{{Eisenstein} {et~al.}(2011){Eisenstein}, {Weinberg}, {Agol},
  {Aihara}, {Allende Prieto}, {Anderson}, {Arns}, {Aubourg}, {Bailey},
  {Balbinot}, \& et~al.}]{2011AJ....142...72E}
{Eisenstein}, D.~J., {Weinberg}, D.~H., {Agol}, E., {et~al.} 2011, \aj, 142, 72

\bibitem[{{Ellingson} {et~al.}(2001){Ellingson}, {Lin}, {Yee}, \&
  {Carlberg}}]{2001ApJ...547..609E}
{Ellingson}, E., {Lin}, H., {Yee}, H.~K.~C., \& {Carlberg}, R.~G. 2001, \apj,
  547, 609

\bibitem[{{Fang} {et~al.}(2010){Fang}, {Buote}, {Humphrey}, {Canizares},
  {Zappacosta}, {Maiolino}, {Tagliaferri}, \&
  {Gastaldello}}]{2010ApJ...714.1715F}
{Fang}, T., {Buote}, D.~A., {Humphrey}, P.~J., {et~al.} 2010, \apj, 714, 1715

\bibitem[{{Fioc} \& {Rocca-Volmerange}(1997)}]{1997A&A...326..950F}
{Fioc}, M., \& {Rocca-Volmerange}, B. 1997, \aap, 326, 950

\bibitem[{{Fraser-McKelvie} {et~al.}(2011){Fraser-McKelvie}, {Pimbblet}, \&
  {Lazendic}}]{2011MNRAS.415.1961F}
{Fraser-McKelvie}, A., {Pimbblet}, K.~A., \& {Lazendic}, J.~S. 2011, \mnras,
  415, 1961

\bibitem[{{Geller} \& {Huchra}(1989)}]{1989Sci...246..897G}
{Geller}, M.~J., \& {Huchra}, J.~P. 1989, Science, 246, 897

\bibitem[{{Guo} {et~al.}(2011){Guo}, {White}, {Boylan-Kolchin}, {De Lucia},
  {Kauffmann}, {Lemson}, {Li}, {Springel}, \& {Weinmann}}]{2011MNRAS.413..101G}
{Guo}, Q., {White}, S., {Boylan-Kolchin}, M., {et~al.} 2011, \mnras, 413, 101

\bibitem[{{Hansen} {et~al.}(2009){Hansen}, {Sheldon}, {Wechsler}, \&
  {Koester}}]{2009ApJ...699.1333H}
{Hansen}, S.~M., {Sheldon}, E.~S., {Wechsler}, R.~H., \& {Koester}, B.~P. 2009,
  \apj, 699, 1333

\bibitem[{{Hao} {et~al.}(2010){Hao}, {McKay}, {Koester}, {Rykoff}, {Rozo},
  {Annis}, {Wechsler}, {Evrard}, {Siegel}, {Becker}, {Busha}, {Gerdes},
  {Johnston}, \& {Sheldon}}]{2010ApJS..191..254H}
{Hao}, J., {McKay}, T.~A., {Koester}, B.~P., {et~al.} 2010, \apjs, 191, 254

\bibitem[{{Henriques} {et~al.}(2012){Henriques}, {White}, {Lemson}, {Thomas},
  {Guo}, {Marleau}, \& {Overzier}}]{2012MNRAS.421.2904H}
{Henriques}, B.~M.~B., {White}, S.~D.~M., {Lemson}, G., {et~al.} 2012, \mnras,
  421, 2904

\bibitem[{{Hoyle} {et~al.}(2005){Hoyle}, {Rojas}, {Vogeley}, \&
  {Brinkmann}}]{2005ApJ...620..618H}
{Hoyle}, F., {Rojas}, R.~R., {Vogeley}, M.~S., \& {Brinkmann}, J. 2005, \apj,
  620, 618

\bibitem[{{Hoyle} {et~al.}(2012){Hoyle}, {Vogeley}, \&
  {Pan}}]{2012MNRAS.426.3041H}
{Hoyle}, F., {Vogeley}, M.~S., \& {Pan}, D. 2012, \mnras, 426, 3041

\bibitem[{{J{\~o}eveer} {et~al.}(1978){J{\~o}eveer}, {Einasto}, \&
  {Tago}}]{1978MNRAS.185..357J}
{J{\~o}eveer}, M., {Einasto}, J., \& {Tago}, E. 1978, \mnras, 185, 357

\bibitem[{{Jauzac} {et~al.}(2012){Jauzac}, {Jullo}, {Kneib}, {Ebeling},
  {Leauthaud}, {Ma}, {Limousin}, {Massey}, \& {Richard}}]{2012MNRAS.426.3369J}
{Jauzac}, M., {Jullo}, E., {Kneib}, J.-P., {et~al.} 2012, \mnras, 426, 3369

\bibitem[{{Kauffmann}(1995)}]{1995MNRAS.274..153K}
{Kauffmann}, G. 1995, \mnras, 274, 153

\bibitem[{{Klypin} \& {Shandarin}(1983)}]{1983MNRAS.204..891K}
{Klypin}, A.~A., \& {Shandarin}, S.~F. 1983, \mnras, 204, 891

\bibitem[{{Koester} {et~al.}(2007){Koester}, {McKay}, {Annis}, {Wechsler},
  {Evrard}, {Bleem}, {Becker}, {Johnston}, {Sheldon}, {Nichol}, {Miller},
  {Scranton}, {Bahcall}, {Barentine}, {Brewington}, {Brinkmann}, {Harvanek},
  {Kleinman}, {Krzesinski}, {Long}, {Nitta}, {Schneider}, {Sneddin}, {Voges},
  \& {York}}]{2007ApJ...660..239K}
{Koester}, B.~P., {McKay}, T.~A., {Annis}, J., {et~al.} 2007, \apj, 660, 239

\bibitem[{{Li} {et~al.}(2009){Li}, {Yee}, \& {Ellingson}}]{2009ApJ...698...83L}
{Li}, I.~H., {Yee}, H.~K.~C., \& {Ellingson}, E. 2009, \apj, 698, 83

\bibitem[{{Li} {et~al.}(2012){Li}, {Yee}, {Hsieh}, \&
  {Gladders}}]{2012ApJ...749..150L}
{Li}, I.~H., {Yee}, H.~K.~C., {Hsieh}, B.~C., \& {Gladders}, M. 2012, \apj,
  749, 150

\bibitem[{{Malarecki} {et~al.}(2013){Malarecki}, {Staveley-Smith}, {Saripalli},
  {Subrahmanyan}, {Jones}, {Duffy}, \& {Rioja}}]{2013MNRAS.tmp.1162M}
{Malarecki}, J.~M., {Staveley-Smith}, L., {Saripalli}, L., {et~al.} 2013,
  \mnras

\bibitem[{{Margoniner} {et~al.}(2001){Margoniner}, {de Carvalho}, {Gal}, \&
  {Djorgovski}}]{2001ApJ...548L.143M}
{Margoniner}, V.~E., {de Carvalho}, R.~R., {Gal}, R.~R., \& {Djorgovski}, S.~G.
  2001, \apjl, 548, L143

\bibitem[{{Markwardt}(2009)}]{2009ASPC..411..251M}
{Markwardt}, C.~B. 2009, in Astronomical Society of the Pacific Conference
  Series, Vol. 411, Astronomical Data Analysis Software and Systems XVIII, ed.
  D.~A. {Bohlender}, D.~{Durand}, \& P.~{Dowler}, 251

\bibitem[{{McGee} {et~al.}(2009){McGee}, {Balogh}, {Bower}, {Font}, \&
  {McCarthy}}]{2009MNRAS.400..937M}
{McGee}, S.~L., {Balogh}, M.~L., {Bower}, R.~G., {Font}, A.~S., \& {McCarthy},
  I.~G. 2009, \mnras, 400, 937

\bibitem[{{Miller} {et~al.}(2005){Miller}, {Nichol}, {Reichart}, {Wechsler},
  {Evrard}, {Annis}, {McKay}, {Bahcall}, {Bernardi}, {Boehringer}, {Connolly},
  {Goto}, {Kniazev}, {Lamb}, {Postman}, {Schneider}, {Sheth}, \&
  {Voges}}]{2005AJ....130..968M}
{Miller}, C.~J., {Nichol}, R.~C., {Reichart}, D., {et~al.} 2005, \aj, 130, 968

\bibitem[{{Montero-Dorta} \& {Prada}(2009)}]{2009MNRAS.399.1106M}
{Montero-Dorta}, A.~D., \& {Prada}, F. 2009, \mnras, 399, 1106

\bibitem[{{Peng} {et~al.}(2010){Peng}, {Lilly}, {Kova{\v c}}, {Bolzonella},
  {Pozzetti}, {Renzini}, {Zamorani}, {Ilbert}, {Knobel}, {Iovino}, {Maier},
  {Cucciati}, {Tasca}, {Carollo}, {Silverman}, {Kampczyk}, {de Ravel},
  {Sanders}, {Scoville}, {Contini}, {Mainieri}, {Scodeggio}, {Kneib}, {Le
  F{\`e}vre}, {Bardelli}, {Bongiorno}, {Caputi}, {Coppa}, {de la Torre},
  {Franzetti}, {Garilli}, {Lamareille}, {Le Borgne}, {Le Brun}, {Mignoli},
  {Perez Montero}, {Pello}, {Ricciardelli}, {Tanaka}, {Tresse}, {Vergani},
  {Welikala}, {Zucca}, {Oesch}, {Abbas}, {Barnes}, {Bordoloi}, {Bottini},
  {Cappi}, {Cassata}, {Cimatti}, {Fumana}, {Hasinger}, {Koekemoer},
  {Leauthaud}, {Maccagni}, {Marinoni}, {McCracken}, {Memeo}, {Meneux}, {Nair},
  {Porciani}, {Presotto}, \& {Scaramella}}]{2010ApJ...721..193P}
{Peng}, Y.-j., {Lilly}, S.~J., {Kova{\v c}}, K., {et~al.} 2010, \apj, 721, 193

\bibitem[{{Pimbblet} {et~al.}(2004){Pimbblet}, {Drinkwater}, \&
  {Hawkrigg}}]{2004MNRAS.354L..61P}
{Pimbblet}, K.~A., {Drinkwater}, M.~J., \& {Hawkrigg}, M.~C. 2004, \mnras, 354,
  L61

\bibitem[{{Popesso} {et~al.}(2005){Popesso}, {B{\"o}hringer}, {Romaniello}, \&
  {Voges}}]{2005A&A...433..415P}
{Popesso}, P., {B{\"o}hringer}, H., {Romaniello}, M., \& {Voges}, W. 2005,
  \aap, 433, 415

\bibitem[{{Porter} {et~al.}(2008){Porter}, {Raychaudhury}, {Pimbblet}, \&
  {Drinkwater}}]{2008MNRAS.388.1152P}
{Porter}, S.~C., {Raychaudhury}, S., {Pimbblet}, K.~A., \& {Drinkwater}, M.~J.
  2008, \mnras, 388, 1152

\bibitem[{{Roche} {et~al.}(2009){Roche}, {Bernardi}, \&
  {Hyde}}]{2009MNRAS.398.1549R}
{Roche}, N., {Bernardi}, M., \& {Hyde}, J. 2009, \mnras, 398, 1549

\bibitem[{{Rykoff} {et~al.}(2013){Rykoff}, {Rozo}, {Busha}, {Cunha},
  {Finoguenov}, {Evrard}, {Hao}, {Koester}, {Leauthaud}, {Nord}, {Pierre},
  {Reddick}, {Sadibekova}, {Sheldon}, \& {Wechsler}}]{2013arXiv1303.3562R}
{Rykoff}, E.~S., {Rozo}, E., {Busha}, M.~T., {et~al.} 2013, ArXiv e-prints

\bibitem[{{Schechter}(1976)}]{1976ApJ...203..297S}
{Schechter}, P. 1976, \apj, 203, 297

\bibitem[{{Smail} {et~al.}(1998){Smail}, {Edge}, {Ellis}, \&
  {Blandford}}]{1998MNRAS.293..124S}
{Smail}, I., {Edge}, A.~C., {Ellis}, R.~S., \& {Blandford}, R.~D. 1998, \mnras,
  293, 124

\bibitem[{{Springel} {et~al.}(2005){Springel}, {White}, {Jenkins}, {Frenk},
  {Yoshida}, {Gao}, {Navarro}, {Thacker}, {Croton}, {Helly}, {Peacock}, {Cole},
  {Thomas}, {Couchman}, {Evrard}, {Colberg}, \& {Pearce}}]{2005Natur.435..629S}
{Springel}, V., {White}, S.~D.~M., {Jenkins}, A., {et~al.} 2005, \nat, 435, 629

\bibitem[{{Stoughton} {et~al.}(2002){Stoughton}, {Lupton}, {Bernardi},
  {Blanton}, {Burles}, {Castander}, {Connolly}, {Eisenstein}, {Frieman},
  {Hennessy}, {Hindsley}, {Ivezi{\'c}}, {Kent}, {Kunszt}, {Lee}, {Meiksin},
  {Munn}, {Newberg}, {Nichol}, {Nicinski}, {Pier}, {Richards}, {Richmond},
  {Schlegel}, {Smith}, {Strauss}, {SubbaRao}, {Szalay}, {Thakar}, {Tucker},
  {Vanden Berk}, {Yanny}, {Adelman}, {Anderson}, {Anderson}, {Annis},
  {Bahcall}, {Bakken}, {Bartelmann}, {Bastian}, {Bauer}, {Berman},
  {B{\"o}hringer}, {Boroski}, {Bracker}, {Briegel}, {Briggs}, {Brinkmann},
  {Brunner}, {Carey}, {Carr}, {Chen}, {Christian}, {Colestock}, {Crocker},
  {Csabai}, {Czarapata}, {Dalcanton}, {Davidsen}, {Davis}, {Dehnen},
  {Dodelson}, {Doi}, {Dombeck}, {Donahue}, {Ellman}, {Elms}, {Evans}, {Eyer},
  {Fan}, {Federwitz}, {Friedman}, {Fukugita}, {Gal}, {Gillespie}, {Glazebrook},
  {Gray}, {Grebel}, {Greenawalt}, {Greene}, {Gunn}, {de Haas}, {Haiman},
  {Haldeman}, {Hall}, {Hamabe}, {Hansen}, {Harris}, {Harris}, {Harvanek},
  {Hawley}, {Hayes}, {Heckman}, {Helmi}, {Henden}, {Hogan}, {Hogg}, {Holmgren},
  {Holtzman}, {Huang}, {Hull}, {Ichikawa}, {Ichikawa}, {Johnston}, {Kauffmann},
  {Kim}, {Kimball}, {Kinney}, {Klaene}, {Kleinman}, {Klypin}, {Knapp},
  {Korienek}, {Krolik}, {Kron}, {Krzesi{\'n}ski}, {Lamb}, {Leger},
  {Limmongkol}, {Lindenmeyer}, {Long}, {Loomis}, {Loveday}, {MacKinnon},
  {Mannery}, {Mantsch}, {Margon}, {McGehee}, {McKay}, {McLean}, {Menou},
  {Merelli}, {Mo}, {Monet}, {Nakamura}, {Narayanan}, {Nash}, {Neilsen},
  {Newman}, {Nitta}, {Odenkirchen}, {Okada}, {Okamura}, {Ostriker}, {Owen},
  {Pauls}, {Peoples}, {Peterson}, {Petravick}, {Pope}, {Pordes}, {Postman},
  {Prosapio}, {Quinn}, {Rechenmacher}, {Rivetta}, {Rix}, {Rockosi}, {Rosner},
  {Ruthmansdorfer}, {Sandford}, {Schneider}, {Scranton}, {Sekiguchi}, {Sergey},
  {Sheth}, {Shimasaku}, {Smee}, {Snedden}, {Stebbins}, {Stubbs}, {Szapudi},
  {Szkody}, {Szokoly}, {Tabachnik}, {Tsvetanov}, {Uomoto}, {Vogeley}, {Voges},
  {Waddell}, {Walterbos}, {Wang}, {Watanabe}, {Weinberg}, {White}, {White},
  {Wilhite}, {Wolfe}, {Yasuda}, {York}, {Zehavi}, \&
  {Zheng}}]{2002AJ....123..485S}
{Stoughton}, C., {Lupton}, R.~H., {Bernardi}, M., {et~al.} 2002, \aj, 123, 485

\bibitem[{{Werner} {et~al.}(2008){Werner}, {Finoguenov}, {Kaastra},
  {Simionescu}, {Dietrich}, {Vink}, \& {B{\"o}hringer}}]{2008A&A...482L..29W}
{Werner}, N., {Finoguenov}, A., {Kaastra}, J.~S., {et~al.} 2008, \aap, 482, L29

\bibitem[{{Williams} {et~al.}(2010){Williams}, {Bureau}, \&
  {Cappellari}}]{2010MNRAS.409.1330W}
{Williams}, M.~J., {Bureau}, M., \& {Cappellari}, M. 2010, \mnras, 409, 1330

\bibitem[{{Yang} {et~al.}(2008){Yang}, {Mo}, \& {van den
  Bosch}}]{2008ApJ...676..248Y}
{Yang}, X., {Mo}, H.~J., \& {van den Bosch}, F.~C. 2008, \apj, 676, 248

\bibitem[{{Zeldovich}(1970)}]{1970A&A.....5...84Z}
{Zeldovich}, Y.~B. 1970, \aap, 5, 84

\end{thebibliography}
\appendix

\section{Detection of Filaments in Simulation Data}
\label{section:mock}

As further test of our method as well as for comparison with
simulation data, we run our algorithm on the Millennium simulation
\citep{2005Natur.435..629S}, a large volume, high resolution N-body
simulation under $\Lambda$CDM cosmology. This simulation includes
information of dark matter structures, i.e., Friends-of-Friends (FOF)
groups, and dark matter substructures. It also tracks the merging
history of dark matter structures and substructures for galaxy
formation simulation. For our purpose, we use the all sky lightcone
realization of this simulation constructed with methods fully
described in \cite{2012MNRAS.421.2904H} and
\cite{2005MNRAS.360..159B}, under the galaxy formation simulation of
\cite{2011MNRAS.413..101G}.

Due to differences between these simulation data and SDSS
observational data, we make a few adjustments of our methods before
running our algorithm. First, instead of using gmBCG clusters to build
the cluster pair catalog, we search for pairs of central BCGs in
cluster-sized FOF groups with $M_{200}$ (the mass within the radius
where the enclosed average density is 200 times the critical density
of the simulation) larger than $10^{14}M_{\sun}$. We make this
adjustment because running the gmBCG cluster finder on these
simulations does not yield satisfying purity and completeness,
possibly due to the lack of clear color bimodality in the simulated
galaxy population.  Secondly, the simulation galaxy catalog has a
apparent magnitude limit in i-band at
$i=21.0\,\mathrm{mag}$. Therefore, when making the brightness cut to
ensure fair comparison of the galaxy population at different redshifts
(as described in Section~\ref{section:brightness_cut}), the luminosity
limit is adjusted to $-25.3\le M_i \le -21.9$. This luminosity limit is
above the $i=21.0\,\mathrm{mag}$ apparent magnitude limit at
$z=0.38$. For a fair comparison, we also rerun our codes with SDSS
data under the new luminosity cut. When comparing results from
simulation data and SDSS data in this section, all the results
presented are under this new luminosity cut.

\begin{figure*}
  \begin{center}
    \plotone{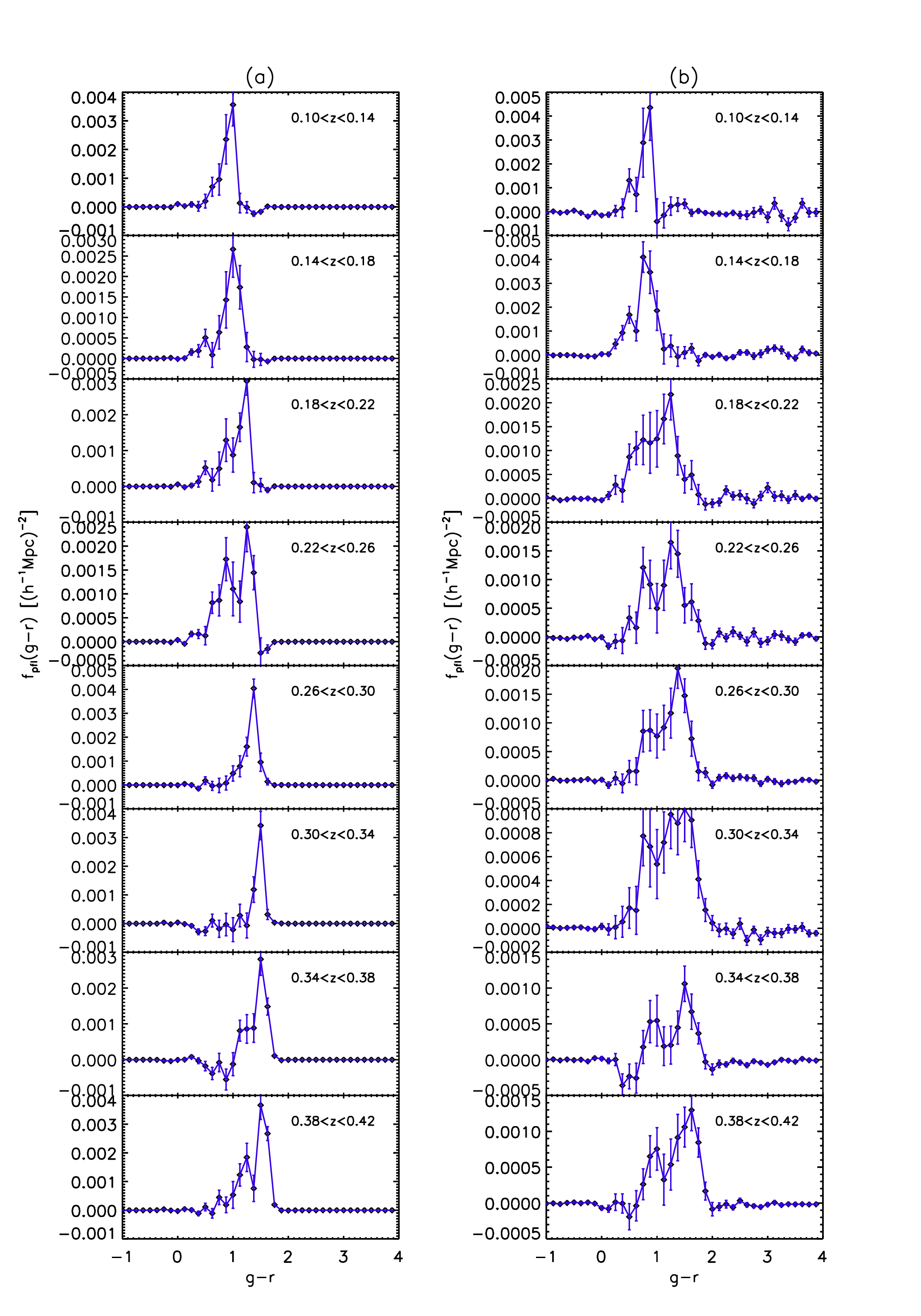}
    \caption{$g-r$ distributions of filaments at different redshift bins
      in simulation (a) and SDSS (b). Comparing to SDSS data, the
      simulation is over-populated with red galaxies. Note that the
      galaxy population is incomplete at $0.38<z<0.42$ (See discussion
      in Section~\ref{section:brightness_cut}) and also (b) differs
      from Figure~\ref{fig:gr100} (b) because of the different
      luminosity cut applied. See the electronic edition of the
      journal for a color version of this figure.}
    \label{fig:mock_gr}
  \end{center}
\end{figure*}

\begin{figure*}
\begin{center}
  \includegraphics[width=1.0\textwidth]{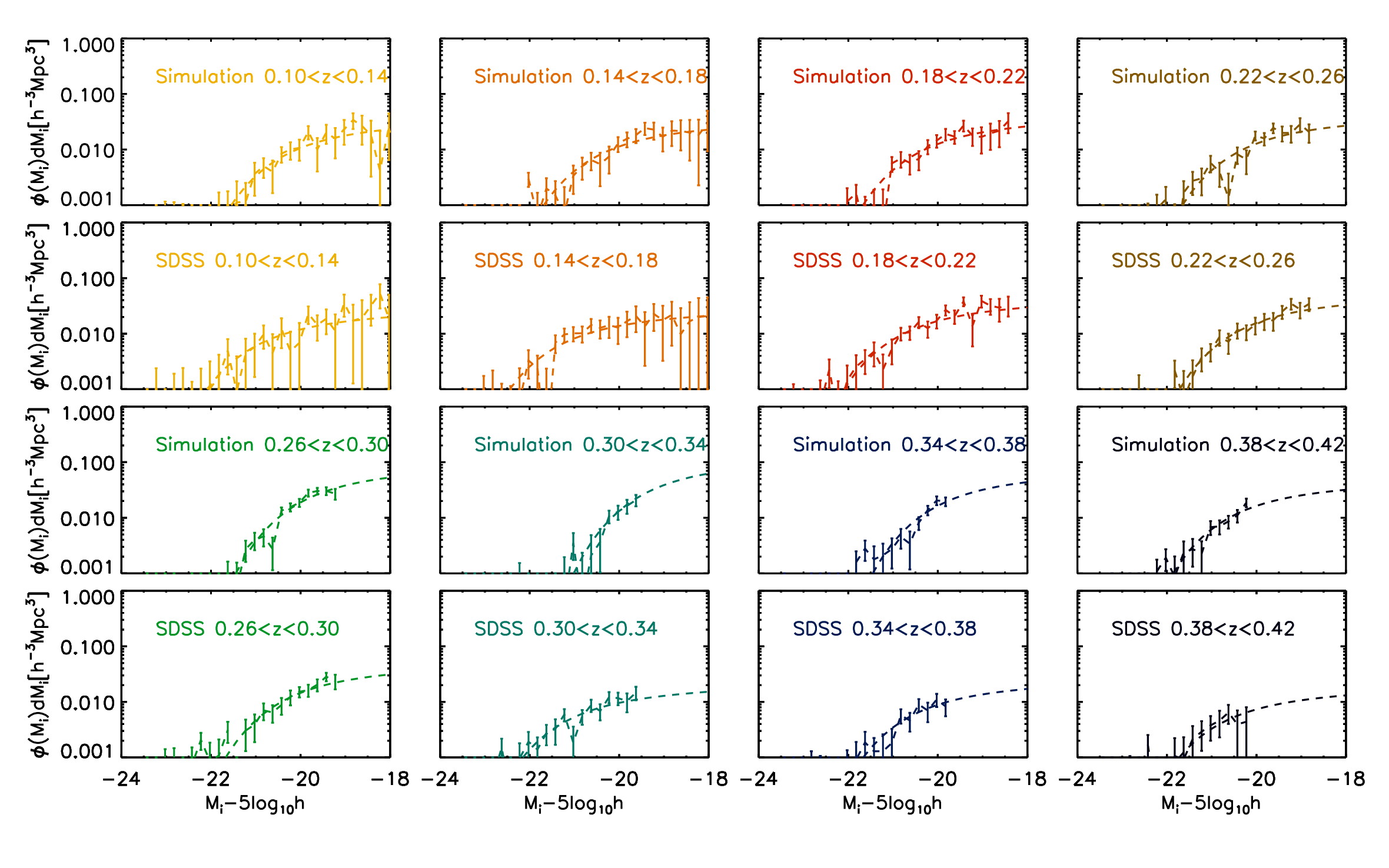}
  \caption{Luminosity distributions of filament galaxies in simulation
    (1st and 3rd rows) and SDSS (2nd and 4th rows) and their Schechter
    function fits (dotted lines).The Schechter function fit has a
    fixed faint end slope, and the rest of the fitting parameters are
    listed in Table~\ref{tbl:mock_lf}. The characteristic magnitude of
    the fitted Schechter function is brighter in simulation than in
    SDSS, indicating overpopulation of bright galaxies in the simulation. See the electronic edition of the
      journal for a color version of this figure.}
\label{fig:mock_mag}
\end{center}
\end{figure*}

In Figure~\ref{fig:mock_gr}, we show the color distributions of
filament galaxies in the Millennium simulations compared to those in
the SDSS data. As noted by \cite{2011MNRAS.413..101G}, the simulation
seems to contain too many red sequence galaxies in filaments, and the
fraction of the blue galaxies does not match that seen in SDSS.

In addition, filament galaxies in the simulation appear to be fainter than
observed in SDSS. In Figure~\ref{fig:mock_mag}, we show the luminosity
distributions of filament galaxies in simulation and SDSS, and fit
them to a Schechter function with the faint end slope fixed to be
$\alpha=-1.2$, as having been discussed in
Section~\ref{section:luminosity}. The fit parameters of the
distributions are listed in Table~\ref{tbl:mock_lf}. Filament galaxies
in simulation appear to have a fainter characteristic magnitude in
i-band than in SDSS. Possibly because of the magnitude cut $-25.3\le M_i \le -21.9$ being too shallow, the Schechter function fittings yield smaller estimations of $\phi^*$ than Section~\ref{section:luminosity}.

\begin{table*}
\begin{center}
\caption{Fitted Luminosity Function Parameters in Simulation and SDSS \label{tbl:mock_lf}}
\begin{tabular}{ccccccccr}
\tableline\tableline
& $z$ & $M_i$ cut & $\phi_1^*[(h^{3}\mathrm{Mpc}^{-3}]$ (Simulation)& $M_{i,1}^*-5\log_{10}h$ $[\mathrm{mag}]$ (Simulation)& $\phi_2^*[(h^{3}\mathrm{Mpc}^{-3}]$ (SDSS)& $M_{i,2}^*-5\log_{10}h$ $[\mathrm{mag}]$ (SDSS)\\
\tableline
&[0.10, 0.14] &$-25.3\le M_i \le-19.5$&$0.12 \pm 0.05$& $-20.39 \pm 0.23$& $0.13 \pm 0.09 $&$-20.63 \pm 0.91$&\\
&[0.14, 0.18] &$-25.3\le M_i \le-20.1$&$0.11 \pm 0.05$& $-20.60 \pm 0.47$& $0.09 \pm 0.05 $&$-21.89 \pm 2.56$&\\
&[0.18, 0.22] &$-25.3\le M_i \le-20.5$&$0.13 \pm 0.04$& $-20.49 \pm 0.28$& $0.15 \pm 0.04 $&$-20.76 \pm 0.23$&\\
&[0.22, 0.26] &$-25.3\le M_i \le-20.9$&$0.14 \pm 0.04$& $-20.43 \pm 0.31$& $0.16 \pm 0.04 $&$-20.43 \pm 0.19$&\\
&[0.26, 0.30] &$-25.3\le M_i \le-21.3$&$0.28 \pm 0.05$& $-20.00 \pm 0.14$& $0.18 \pm 0.04 $&$-20.29 \pm 0.31$&\\
&[0.30, 0.34] &$-25.3\le M_i \le-21.7$&$0.42 \pm 0.18$& $-19.48 \pm 0.31$& $0.07 \pm 0.02 $&$-21.17 \pm 0.30$&\\
&[0.34, 0.38] &$-25.3\le M_i \le-21.9$&$0.27 \pm 0.13$& $-20.07 \pm 0.38$& $0.08 \pm 0.02 $&$-20.64 \pm 0.30$&\\
&[0.38, 0.42] &$-25.3\le M_i \le-22.3$&$0.19 \pm 0.10$& $-20.56 \pm 0.46$& $0.06 \pm 0.03 $&$-20.88 \pm 0.44$&\\
\tableline
\end{tabular}
\end{center}
This table lists the fitted Schechter function parameters of filament luminosity distributions in simulation and SDSS data. $\phi_{1}^*$ and $\phi_{2}^*$ are the normalization factors of the Schechter function fitting in simulation and SDSS perspectively, and $M_{i,1}^*$ and $M_{i,2}^*$ are the characteristic magnitude of the fitting in simulation and SDSS . The errors of the parameters are evaluated through fitting 40 resamplings of the galaxy sets being stacked at each redshift bin.
\end{table*}

\end{document}